\documentclass[fleqn,usenatbib]{mnras}

\usepackage[T1]{fontenc}
\usepackage{ae,aecompl}

\usepackage{graphicx}	
\usepackage{amsmath}	
\usepackage{amssymb}	

\usepackage{newtxtext,newtxmath}

\newcommand{\Nbody}{$N$-body }
\setcitestyle{aysep={},citesep={,}} 
\newcommand{\omegavec}{\boldsymbol{\omega}}
\newcommand{\phiomega}{\phi_{\omega}}
\newcommand{\thetaomega}{\theta_{\omega}}
\newcommand{\thetaomegai}{\theta_{\rm \omega,i}}
\newcommand{\qrot}{Q_{\rm rot}}
\newcommand{\qran}{Q_{\rm ran}}
\newcommand{\rL}{r_{\rm L}}
\newcommand{\rJ}{r_{\rm J}}
\newcommand{\rh}{r_{\rm h}}
\newcommand{\tff}{t_{\rm ff}}
\newcommand{\trhi}{t_{\rm rh,i}}

\title[Early evolution of rotating star clusters]{Early dynamical evolution of rotating star clusters in a tidal field}

\author[M. Tiongco, E. Vesperini and A.~L. Varri]{
Maria A. Tiongco,$^{1,2}$\thanks{E-mail: maria.tiongco@colorado.edu}
Enrico Vesperini,$^{2}$
and Anna Lisa Varri$^{3,4}$
\\
$^{1}$JILA and Department of Astrophysical and Planetary Sciences, University of Colorado, Boulder, CO 80309, USA\\
$^{2}$Department of Astronomy, Indiana University, Bloomington, IN 47405, USA\\
$^{3}$Institute for Astronomy, University of Edinburgh, Royal Observatory, Blackford Hill, Edinburgh EH9 3HJ, UK\\
$^{4}$School of Mathematics, University of Edinburgh, Kings Buildings, Edinburgh EH9 3JZ, UK
}

\date{Accepted XXX. Received YYY; in original form ZZZ}

\pubyear{2015}

\begin{document}
\label{firstpage}
\pagerange{\pageref{firstpage}--\pageref{lastpage}}
\maketitle

\begin{abstract}
In order to explore how the early internal rotational properties of star clusters are affected by the external potential of their host galaxies, we have run a suite of $N$-body simulations following the early dynamical evolution and violent relaxation of rotating star clusters embedded in a tidal field.  Our study focuses on models for which the cluster's rotation axis has a generic orientation relative to the torque of the tidal field.  The interaction between the violent relaxation process, angular momentum of the cluster, and the external torque creates a complex kinematic structure within the cluster, most prominently a radial variation in the position of the rotation axis along both the polar and azimuthal directions. We also examine the cluster's velocity dispersion anisotropy and show that the projected anisotropy may be affected by the variation of the rotation axis directions within the cluster; the combination of projection effects and the complex kinematical features may result in the measurement of tangential anisotropy in the cluster's inner regions.  We also characterize the structural properties of our clusters as a function of their initial rotation and virial ratio and find that clusters may develop a triaxial morphology and a radial variation of the minor axis not necessarily aligned with the rotation axis.  Finally, we examine the long-term evolution of these complex kinematic features.
\end{abstract}




\section{Introduction}

Bulk internal rotation of star clusters is an important component of the dynamical picture of these systems.
On the theoretical side, several pioneering studies of the evolution of rotating star clusters have demonstrated how two-body relaxation over time redistributes angular momentum from the inner regions of the cluster outward, and causes stars to escape and carry away angular momentum, lowering the rotation in the cluster \citep[see, e.g.,][]{einsel1999,kim2002,kim2004,kim2008,ernst2007,hong2013,tiongco2017}.  Most of the numerical studies that included the effects of a tidal field explored the geometrically simplest case where the cluster's angular momentum vector is parallel or anti-parallel to the torque of the host galaxy \citep[e.g., see][]{boily1999}, but additional kinematic complexities have been shown to arise when more general configurations are considered \citep{tiongco2018}.

A recent dramatic increase in the number and quality of observational kinematic studies has led to major progress towards a complete kinematic portrait of star clusters. Such studies draw from proper-motion data from \textit{Gaia} \citep[see, e.g,][]{bianchini2018,sollima2019,Vasiliev2019}, HST \citep{bellini2017}, and line-of-sight velocities from spectroscopic studies using Integral Field and/or multi-fiber spectrographs \citep[see, e.g.,][]{fabricius2014,kamann2018,boberg2017,Ferraro2018,Lanzoni2018,Lanzoni2018b,wan2021}.  In general agreement with the theoretical models predicting a decrease of the strength of rotation over time, observational studies are starting to show that globular clusters with longer relaxation times (i.e., dynamically younger clusters) tend to have stronger rotation  \citep{kamann2018,bianchini2018}.  With more detailed kinematic studies combining proper motion data and spectroscopic data to get three-dimensional velocities, we are closer to being able to build a complete picture of the phase space of star clusters \citep[see, e.g.,][]{bellini2017,vitral2021,dalessandro2021b}.

The dynamical origin of the rotation observed in star clusters is still a matter of investigation; a number theoretical studies have shown that angular momentum inherited from the molecular cloud out of which the cluster formed or acquired from external gas accreted during the cluster formation, dynamical interactions among the various subclumps in the primordial cluster, and the effects of  the host galaxy tidal field may all play an important role for the origin of clusters' internal rotation \citep{vesperini2014,mapelli2017,tiongco2017,lahen2020,ballone2021,chen2021}.

In this study, we broaden the classic studies of violent relaxation of isolated systems (e.g., see \citealt{lyndenbell1967,vanalbada1982,aarseth1988} and also \citealt{gott1973,hohl1979,aguilar1990} for rotating systems), to consider such a process in the case of rotating clusters in the external tidal field of their host galaxy and explore the kinematic evolution in the general case in which the cluster's internal angular momentum vector is not aligned with the orbital angular momentum vector. One of the issues we will explore concerns the origin of the radial variation of the position angle of rotation axis observed in some globular clusters \citep{gebhardt2000,bianchini2013,boberg2017,usher2021}.  In \citet{tiongco2018}, we followed the long-term evolution of rotating star clusters in an external tidal field and rotation axes initially misaligned with their orbital angular velocity vectors.  Because rotation of the cluster due to tidal synchronization occurs parallel to orbital angular velocity vector, the cluster may develop a radial variation in the direction of the rotation axis, due to the combination of cluster's intrinsic rotation and rotation that develops from evolution towards tidal locking.  In this study we explore whether a radial variation in the rotation axis' orientation can arise also during the violent relaxation phase.

Many new and upcoming studies are also focusing on the kinematics of young star clusters, including cluster members and stars within the cluster's vicinity that may be former members of the cluster \citep[see, e.g.,][]{kuhn2019,armstrong2020,lim2020,jerabkova2021,dalessandro2021a}.  Because these star clusters are young, the escaped stars may carry some memory of the cluster's early dynamical history.  Motivated by these observational studies, we will also focus our attention on the differences in the structural and kinematical properties between the cluster's inner, outer, and extratidal regions and on how these complexities may appear in observations along generic lines of sight.

This paper is structured as follows: in Section 2, we discuss the setup of our $N$-body simulations, in Section 3, we present the results of our simulations, and in Section 4, we conclude with a summary of our study.

\section{Method and Initial Conditions}
\label{methods}

We have run $N$-body simulations of star clusters on circular orbits about a point-mass galaxy, using the GPU-accelerated $N$-body code, \textsc{nbody6} \citep{aarseth2003,nitadori2012}. The equations of motion are solved in a co-rotating reference frame centred on the cluster \citep{heggie2003}, rotating with angular speed $\Omega$, equal to that of the angular speed of the cluster's orbit.  The $x$-$y$ plane of coordinate system is coincident with the cluster's orbital plane, where the positive $x$-axis of the rotating reference frame always points away from the galactic centre, and the positive $y$-axis points in the direction of the cluster's motion.  The cluster's orbital angular velocity vector is parallel to the positive $z$-axis.

Similar to \citet{tiongco2018,tiongco2019}, the analysis of our results is executed in a non-rotating reference frame centred on the cluster.  We will refer to the angular velocity vector of the cluster, $\omegavec$, and we also define this vector's direction in space using the angles $\phiomega$ and $\thetaomega$: $\phiomega$ is the angle between the projection of $\omegavec$ in the $x$-$y$ plane (the orbital plane) and the positive $x$-axis, measured in a counter-clockwise direction from the positive $x$-axis, and $\thetaomega$ is the angle between $\omegavec$ and the $z$-axis, measured starting from the positive $z$-axis (the axis pointing outwards from the orbital plane).

Our simulations follow the evolution of star clusters initially with some rotation and also undergoing a phase of violent relaxation.  The violent relaxation phase is set up by setting the system's initial virial ratio, $Q=T/|V|$ (where $T$ and $V$ are the kinetic energy and potential energy of the system, respectively) to a value less than 0.5, so that the system is not in virial equilibrium.  We also decompose $Q$ into its rotational and random components, $\qrot$ and $\qran$, where the kinetic energy of each component is from bulk rotation of the system or random motions of the particles, respectively (we adopt the nomenclature of the components from \citealt{gott1973}).  The clusters are initially spherical with uniform density, with random velocities drawn from a Gaussian distribution, and rotation following a solid-body rotational profile.  The initial size of our system is defined by $\rL$, the limiting radius, and we set the ratio of $\rL$ to the Jacobi radius, $\rJ$, to 0.5.   Notable to our study, we also set the initial orientation of the rotation axis (also defined as the direction of $\omegavec$) relative to the tidal field as a free parameter.  

Table \ref{tab:table1} summarizes the initial parameters for our simulations.  We note that  $\qrot$ is given in the non-rotating reference frame.  Also, we do not aim to characterize in detail the evolution of systems that begin with $\thetaomegai$ = 0 or 180, i.e., parallel and anti-parallel with the orbital angular velocity vector, as the evolution of these configurations is straightforward and has been studied by the references mentioned in the introduction.

The simulations are run with 131~072 equal-mass stars, for approximately 18 free-fall times:

\begin{equation}
\tff=\sqrt{\frac{3\pi}{32G\rho}},
\end{equation}

\noindent
\citep[see, e.g.,][]{bt08}, except for simulation `Cold~th45' which we have run until approximately 9 initial half-mass relaxation times:

\begin{equation}
\trhi=\frac{0.138 N^{1/2}r_{\rm h}^{3/2}}{\langle m\rangle^{1/2} G^{1/2} \log(0.11 N)},
\end{equation}

\noindent
\citep[see, e.g.,][]{heggie2003}, i.e., until core-collapse of the system, for a preliminary study illustrating the long-term evolution of these clusters.  For part of our analysis we will also include stars outside of the Jacobi radius, $\rJ = (GM/3\Omega^2)^{1/3}$.  Stars are removed from the simulation when they move beyond a radius equal to two times $\rJ$.

\begin{table}
\caption{Information about the \Nbody simulations presented in this study. The symbol $\thetaomegai$ denotes the initial value of the angle between the direction of the global angular velocity vector and the $z$-axis, measured starting from the $z$-axis (which points outwards from the orbital plane). For further details, see Section \ref{methods}.}
\label{tab:table1}
\begin{tabular}{@{}cccc}
\hline
Model ID & 
$\qrot$ &
$\qran$ &
$\thetaomegai$ (degrees) \\
\hline
Even~th45 &  0.1 & 0.1 & 45 \\
Even~th90 &  " & " & 90 \\
Even~th135 &  " & " & 135 \\

Cold~th45 &  0.1 & 0.05 & 45 \\
Cold~th90 &  " & " & 90 \\
Cold~th135 &  " & " & 135 \\

Slow~th45 &  0.05 & 0.1 & 45 \\
Slow~th90 &  " & " & 90 \\
Slow~th135 &  " & " & 135 \\
\hline
\end{tabular}
\end{table}

\section{Results}

\subsection{General properties and visualization of the violent relaxation process}

\begin{figure*}
	\includegraphics[width=\textwidth]{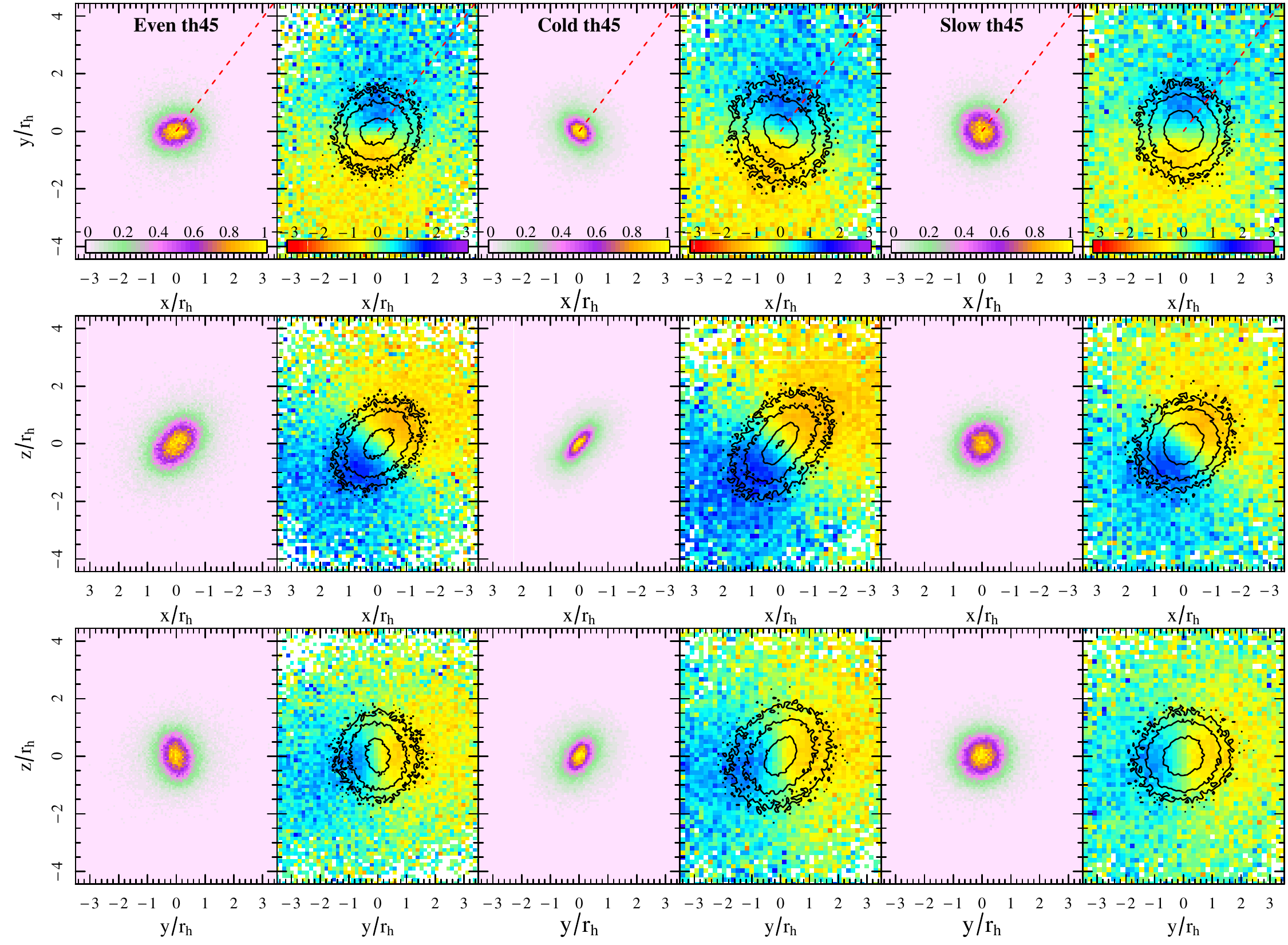}
    \caption{Final snapshot of the Even th45, Cold th45, and Slow th45 simulations (each simulation is featured in a set of 6 panels spanning two columns).  Shown in columns 1, 3, and 5 are the projected number density (normalized to the densest region) along 3 different projections, and in columns 2, 4, and 6, the line-of-sight (los) velocities in the corresponding projections to the projected number density images, with the contours of the projected number density overlaid; the levels of each contour (outermost to innermost) are 10, 25, 125, and 500 stars per pixel.  The los velocities are normalized to $\Omega r_{\rm J,i}$, the speed of co-rotation at the initial Jacobi radius.  The vertical and horizontal axes are normalized to the 3d half-mass radius.  The red dotted line points from the centre of the cluster to the galactic centre.  White pixels in the bottom panels correspond to values either outside the range of the colourbar, or an absence of stars in that pixel.  Only stars inside the Jacobi radius are represented in these images.}
    \label{fig:images1}
\end{figure*}

\begin{figure*}
	\includegraphics[width=\textwidth]{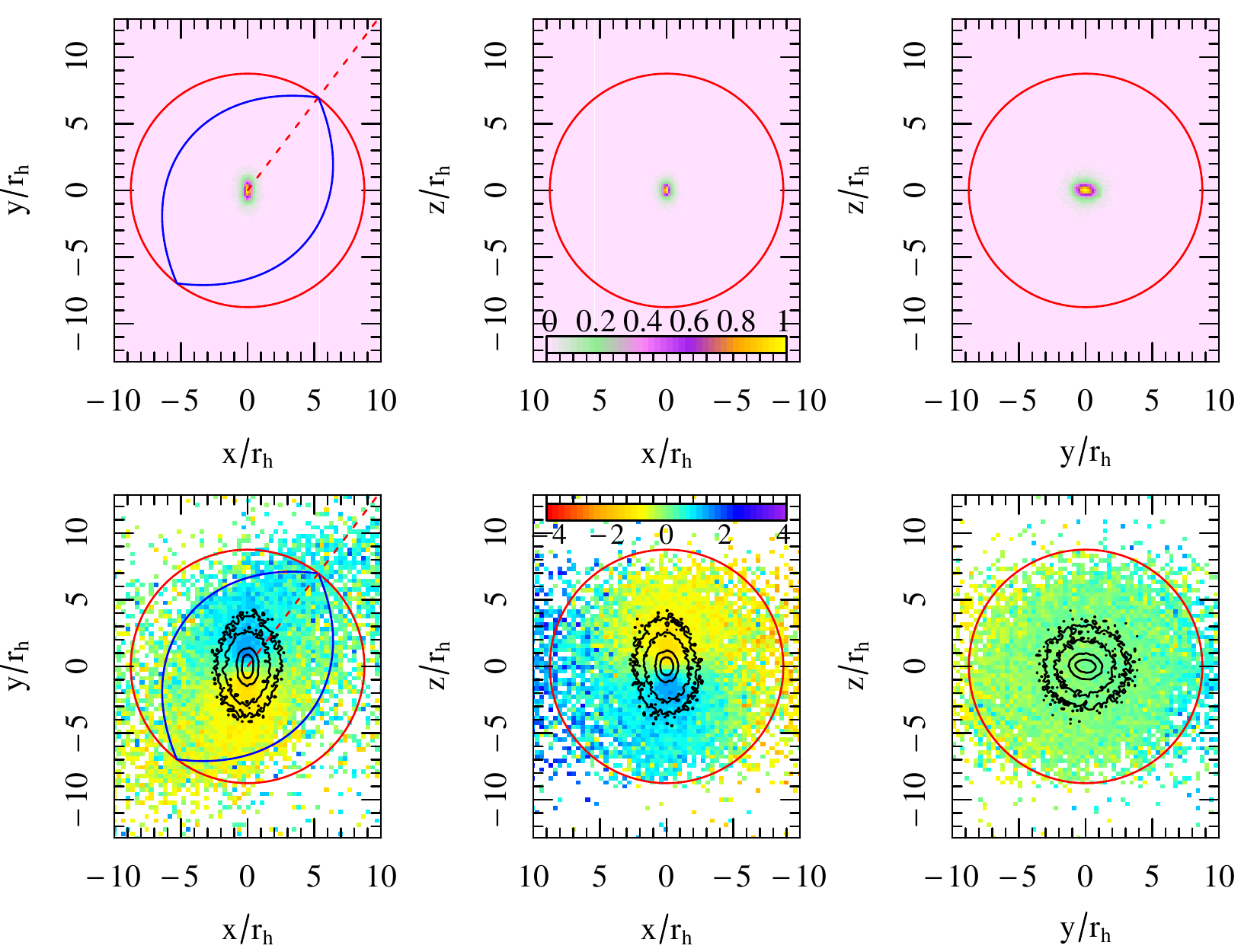}
    \caption{Final snapshot of the Cold th90 simulation, similar to Fig. \ref{fig:images1}, but zoomed out to emphasize the difference in the kinematics between the inner and outer and extratidal regions, e.g., see the bottom middle panel.  The red circle represents the Jacobi radius of the cluster, while the blue lines in the leftmost panels represent a more accurate shape of the Jacobi surface in that projection; the axis ratios of the Jacobi surface are approximately 2/3 for the middle/major axis and 0.5 for the minor/major axis \citep[see, e.g.,][]{heisler1986}.}
    \label{fig:images4}
\end{figure*}

\begin{figure*}
	\includegraphics{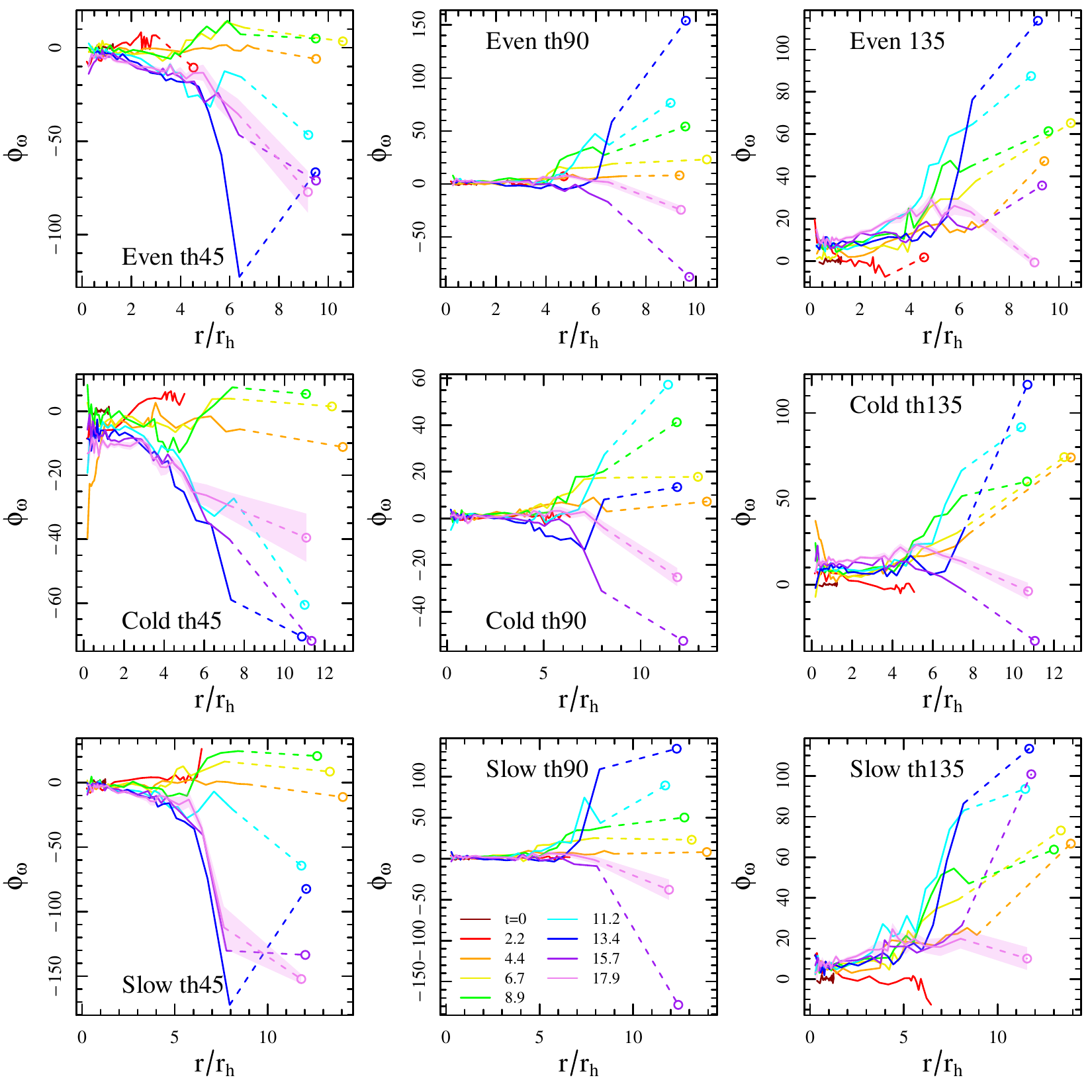}
    \caption{Time evolution of $\phiomega$ (the $\phi$ component of the angular velocity vector, given in degrees) as a function of radius normalized to the half-mass radius, $\rh$, of the cluster at the time specified, shown for all of our models.  The small circles connected with a dashed line are $\phiomega$ calculated from stars outside of the Jacobi radius.  Times in the legend are given in free-fall times, $\tff$. Typical error bands are shown for the final snapshot and are derived using bootstrap sampling of the snapshot. }
    \label{fig:phiom_pro}
\end{figure*}

\begin{figure*}
	\includegraphics{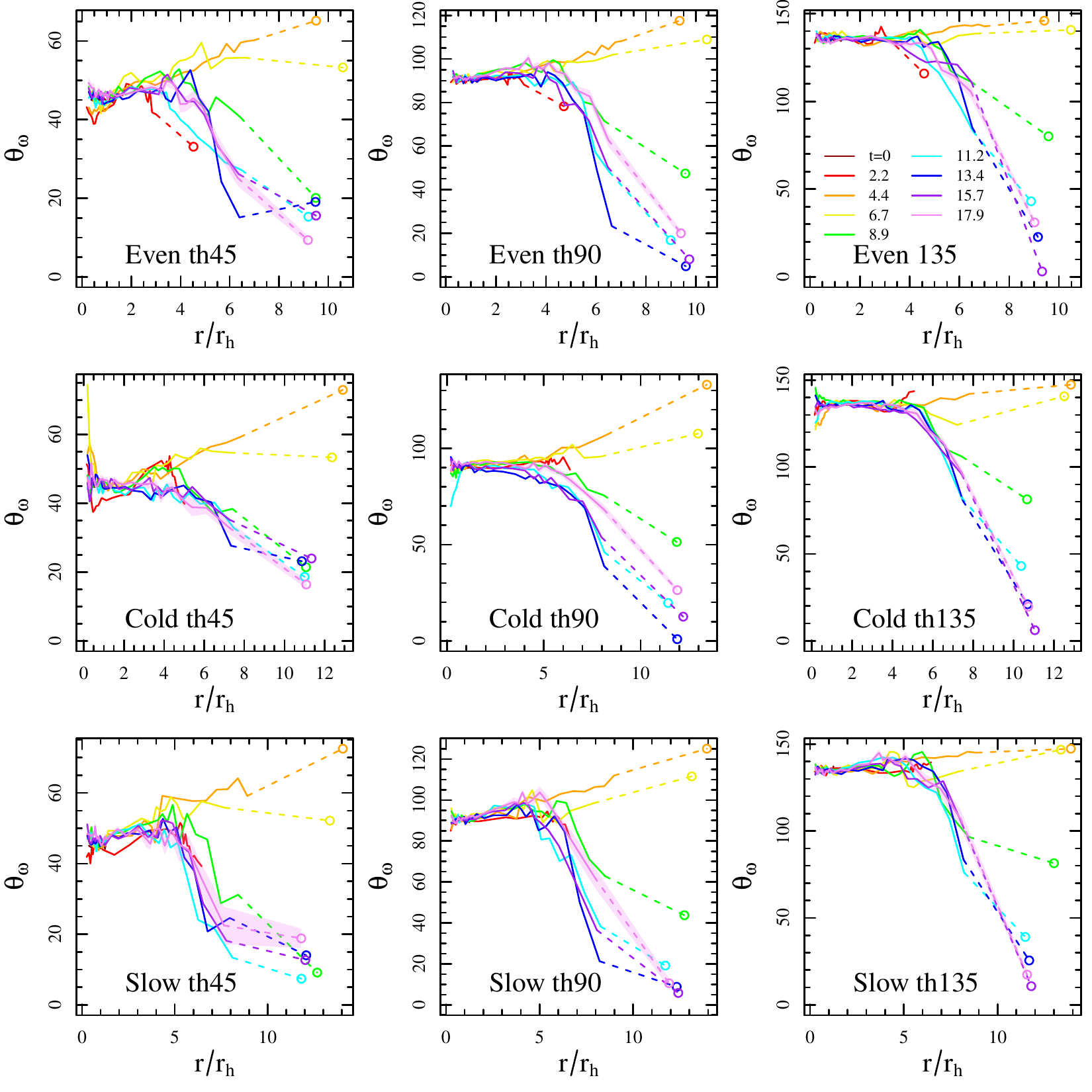}
    \caption{Time evolution of $\thetaomega$ (the $\theta$ component of the angular velocity vector, given in degrees) as a function of radius normalized to the half-mass radius, $\rh$, of the cluster at the time specified, shown for all of our models.  The small circles connected with a dashed line are $\thetaomega$ calculated from stars outside of the Jacobi radius.  Times in the legend are given in free-fall times, $\tff$.  Typical error bands are shown for the final snapshot and are derived using bootstrap sampling of the snapshot.}
    \label{fig:thetaom_pro}
\end{figure*}

\begin{figure*}
	\includegraphics{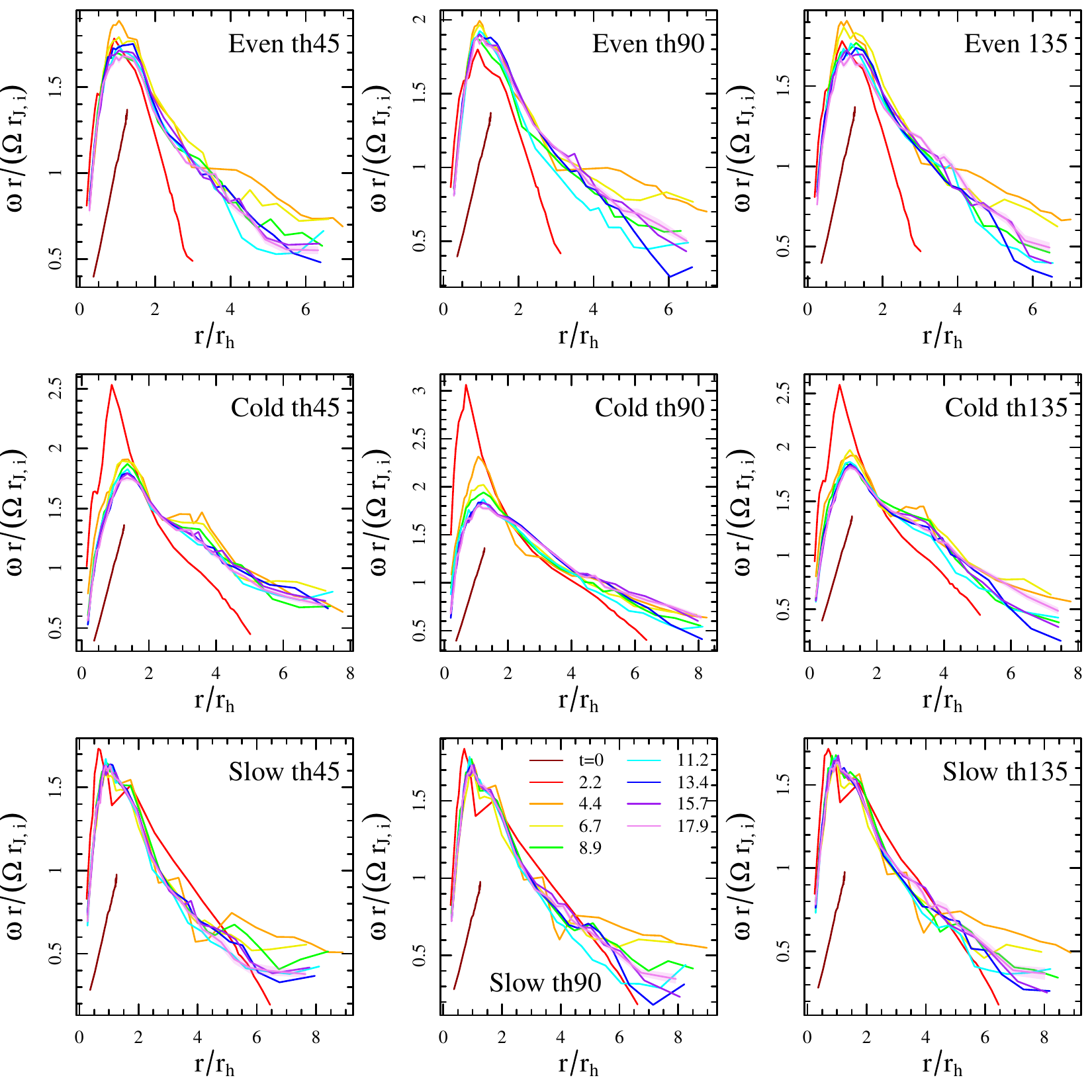}
    \caption{Time evolution of angular velocity vector magnitude multiplied by the radius it was measured at to give a rotational velocity at that radius (radius is normalized to the half-mass radius, $\rh$).  Figs. \ref{fig:phiom_pro} and \ref{fig:thetaom_pro} show the direction of the angular velocity vector, $\omegavec$, and the colors of the lines correspond to the same times in these mentioned figures.  The rotational velocities are also normalized to the velocity of the rotating coordinate system at the initial Jacobi radius.  Typical error bands are shown for the final snapshot and are derived using bootstrap sampling of the snapshot.}
    \label{fig:omegar}
\end{figure*}

\begin{figure*}
	\includegraphics{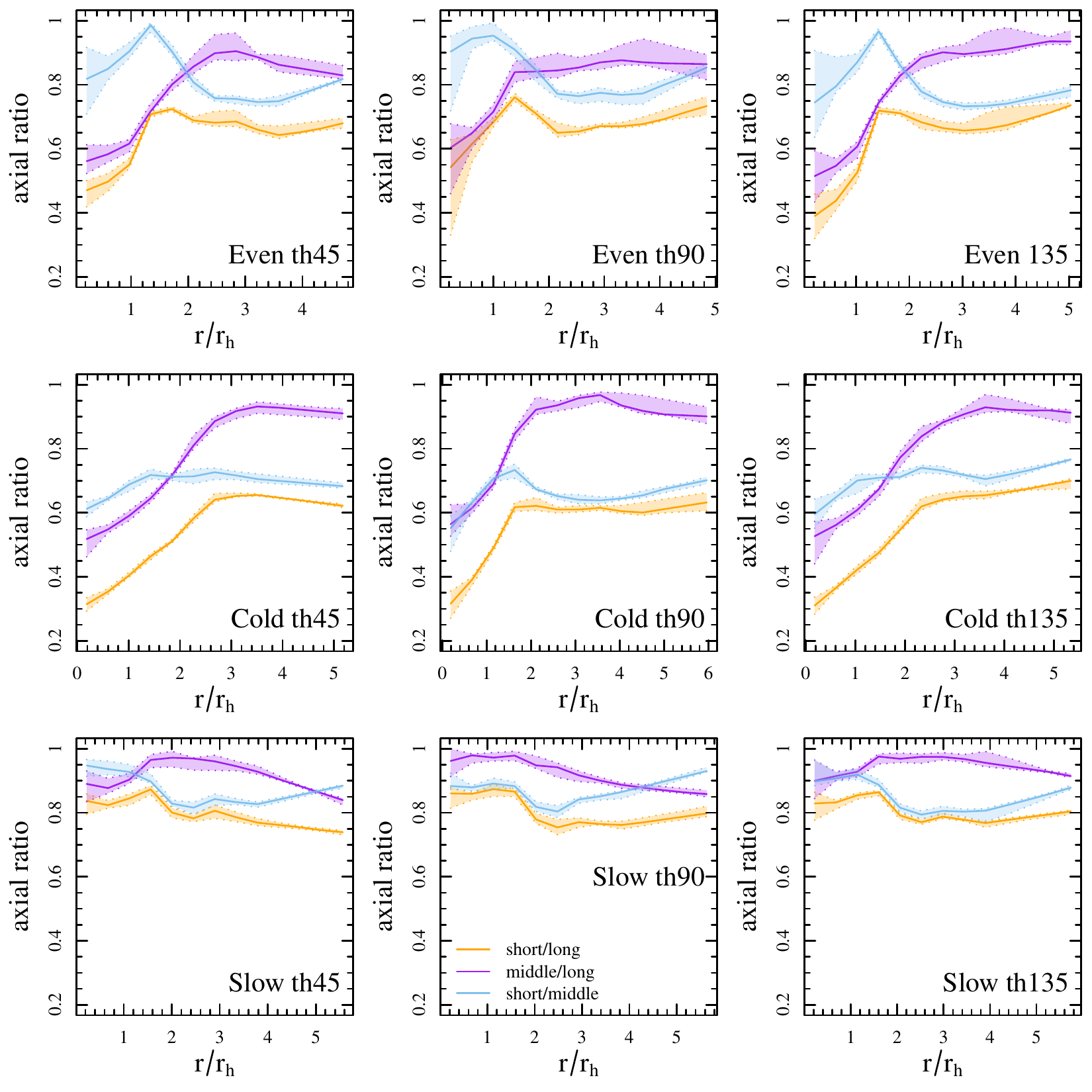}
    \caption{Averaged radial profiles (where radius is normalized to the half-mass radius, $\rh$) of the axial ratios of all of models.  The solid lines show the median profiles of the snapshots between 17.1--17.9$\tff$ and the dashed lines show the maximum and minimum profiles over this timespan.}
    \label{fig:axialratio}
\end{figure*}

\begin{figure*}
    \includegraphics{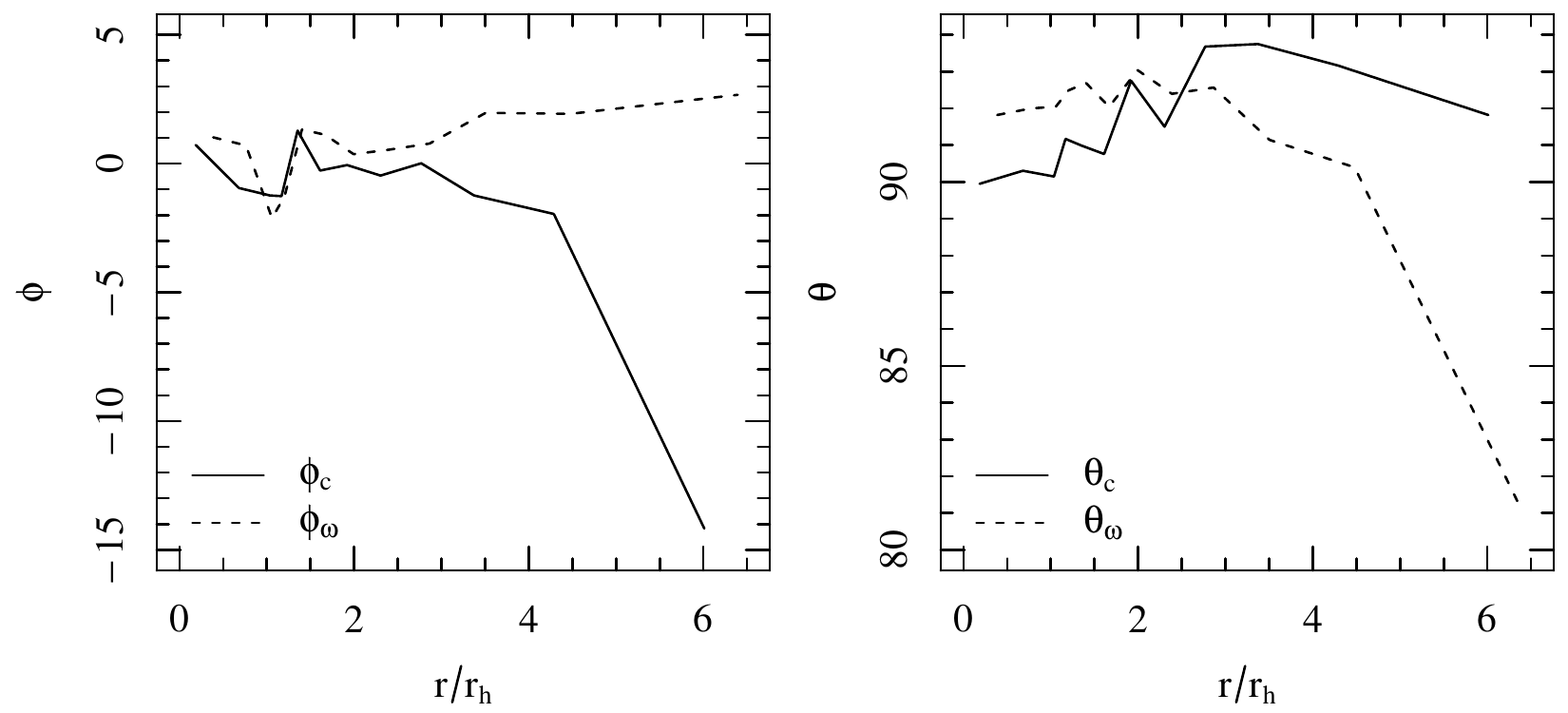}
    \caption{For model `Cold th90', a comparison of the profiles of $\phiomega$ and $\thetaomega$ with $\phi$ and $\theta$ of minor axis ($\phi_{\rm c}$ and $\theta_{\rm c}$, all in degrees) at the end of the simulation.  We find this misalignment between the rotation axis and the minor axis appears in our other models and throughout the timescale of the simulations.}
    \label{fig:morphkin1}
\end{figure*}

\begin{figure*}
	\includegraphics{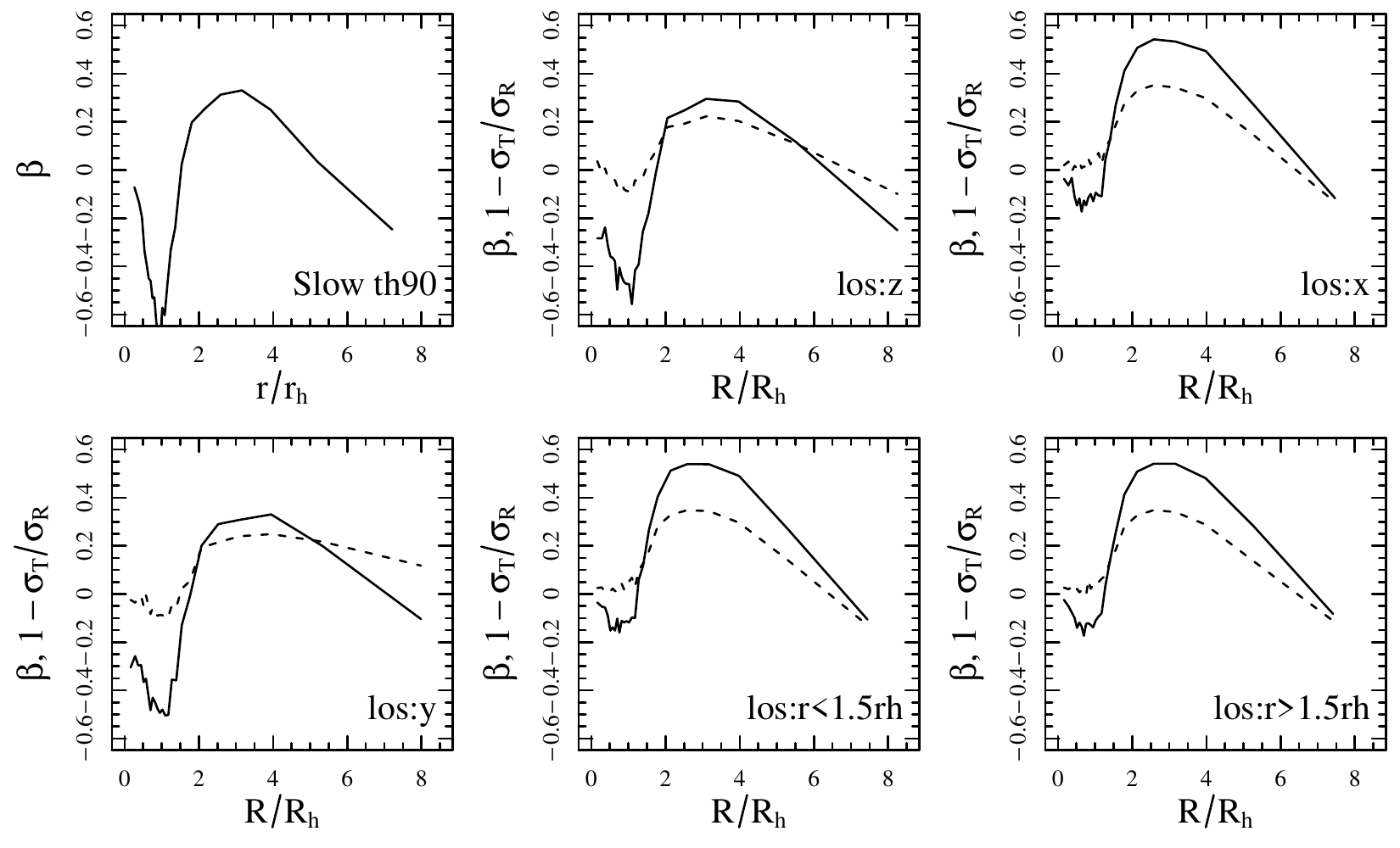}
    \caption{Velocity dispersion anisotropy radial profiles of model `Slow th90' using various binning prescriptions and lines of sight (los) at the end of the simulation.  The top left panel shows the profile of the traditional anisotropy parameter, $\beta$ (see Eq. \ref{eq:beta}), calculated in spherical shells.  In the remaining panels, the solid line represents $\beta$ calculated in cylindrical shells aligning them parallel to the los stated in the legend of the panel.  The bottom-middle and bottom-right panel use a los parallel to the angular velocity vector $\omegavec$ determined from stars within 1.5$\rh$ and greater than 1.5$\rh$, respectively.  The dashed lines represent the projected anisotropy parameter $1-\sigma_{\rm T}/\sigma_{\rm R}$, where the velocity dispersions are calculated from velocities projected onto the plane perpendicular to the los.  $R$ and $R_{\rm h}$ refer to the cylindrical radius and projected half-mass radius.  This model was chosen to show the similarities between the top left panel (`los:x') and the bottom middle panel (`los:r<1.5rh') as this model was initiated with its $\omegavec$ parallel to the $x$-axis.}
    \label{fig:aniso_s90}
\end{figure*}

\begin{figure*}
	\includegraphics{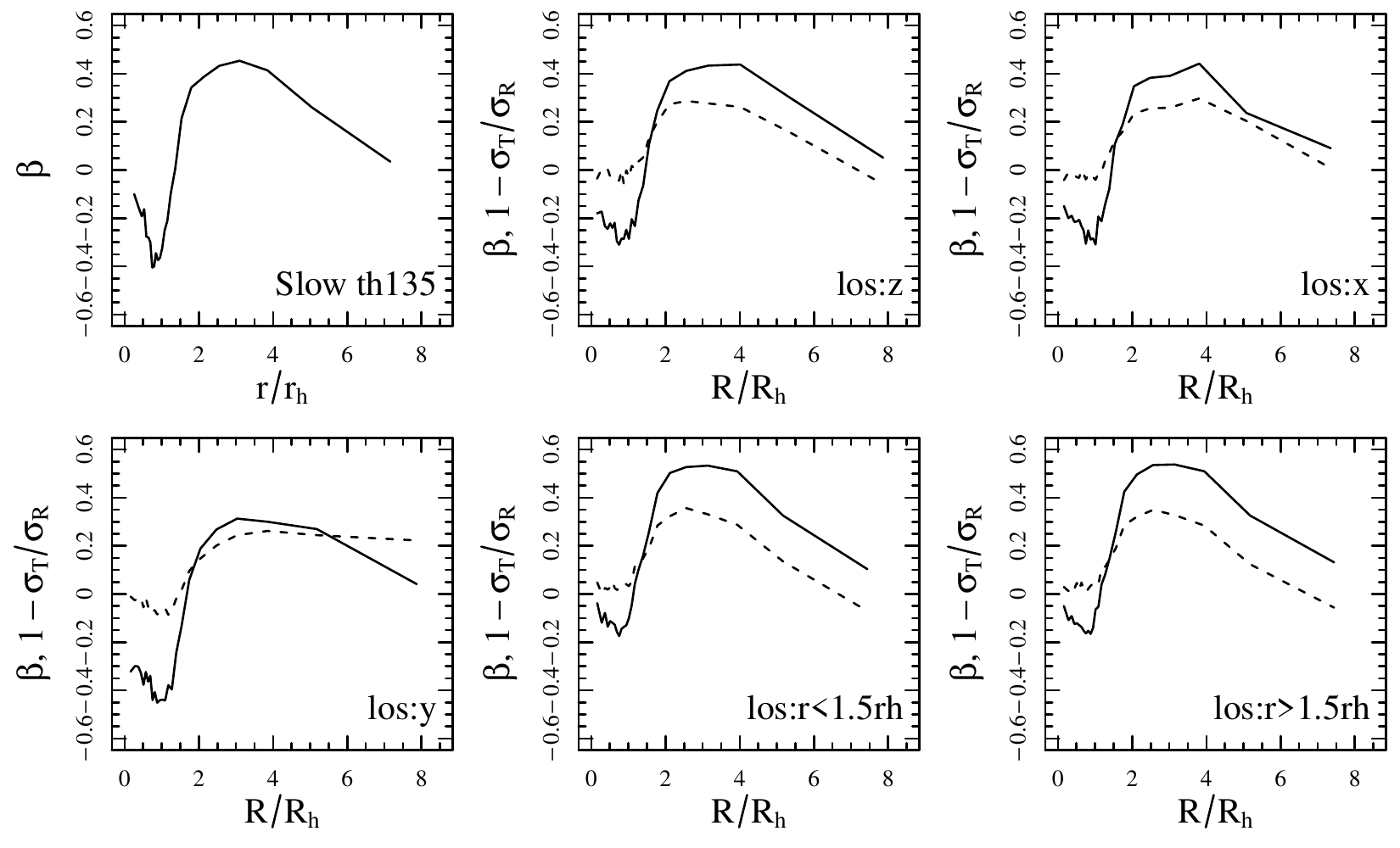}
    \contcaption{Velocity dispersion anisotropy at different lines of sight, different binning prescriptions, and different anisotropy prescriptions (see previous set of panels) for model `Slow 135.'  This model was chosen because its initial $\omegavec$ is not aligned with the $x$-, $y$-, or $z$-axes.}
\end{figure*}

\begin{figure*}
	\includegraphics{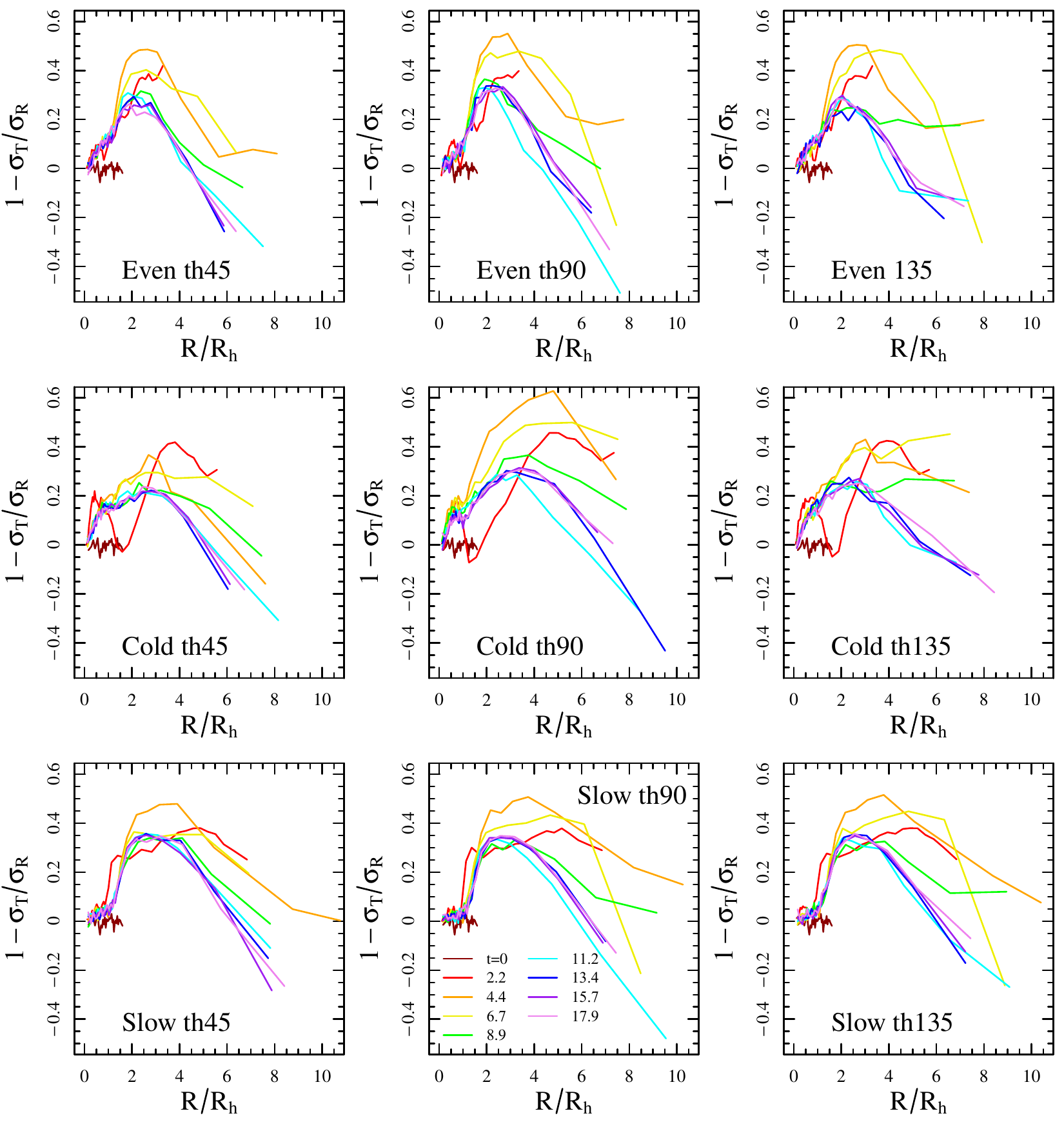}
    \caption{Time evolution of the radial profile of the projected velocity anisotropy using a line-of-sight parallel to the $\omegavec$ of stars within 1.5 $\rh$ to remove as much of the inner tangential bias as possible. $R$ and $R_{\rm h}$ refer to the projected radius and projected half-mass radius.}
    \label{fig:aniall}
\end{figure*}

\begin{figure*}
	\includegraphics{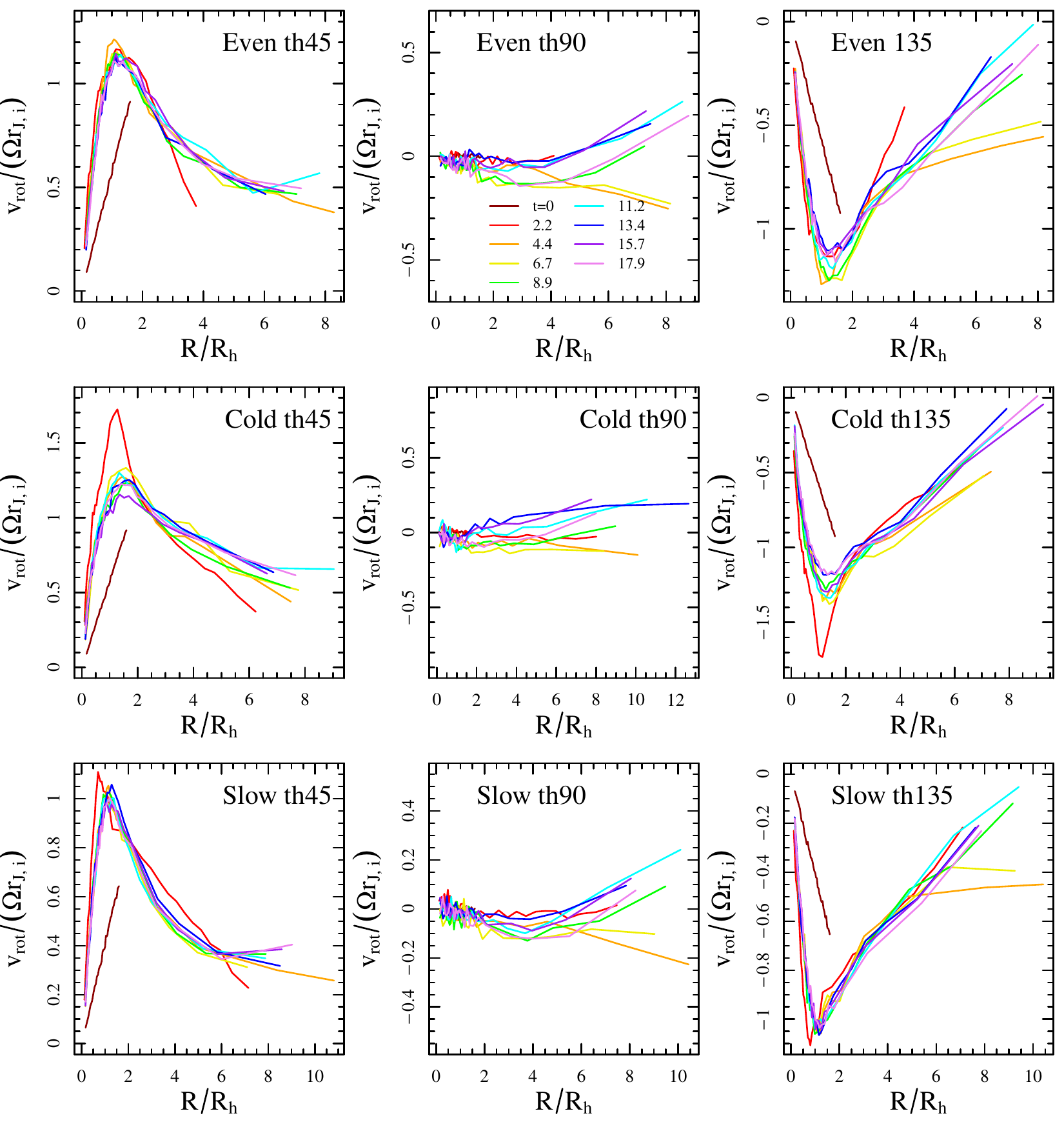}
    \caption{Time evolution of the rotational velocity profile about the $z$-axis in all systems. $R$ and $R_{\rm h}$ refer to the projected radius and projected half-mass radius, respectively.  The rotational velocities are also normalized to the velocity of the rotating coordinate system at the initial Jacobi radius.}
    \label{fig:rotall}
\end{figure*}

\begin{figure*}
	\includegraphics{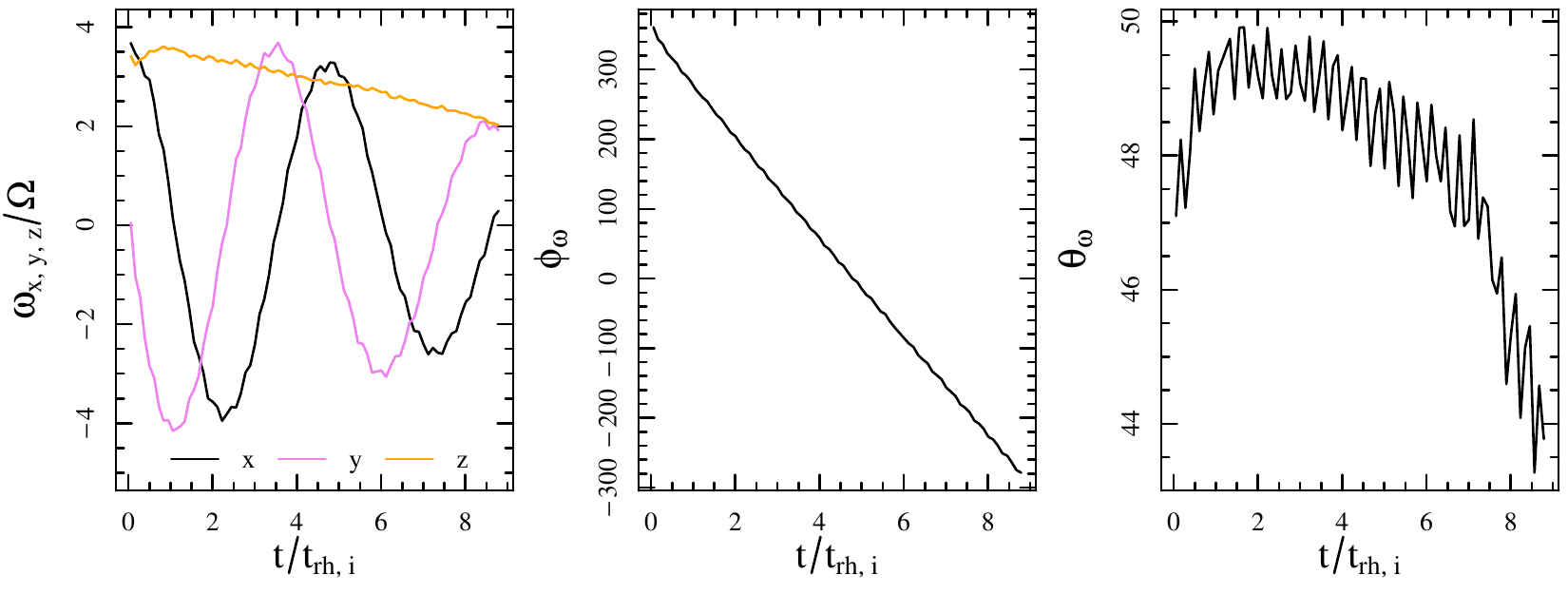}
    \caption{Left panel: time evolution of the Cartesian components of $\omegavec$ calculated on the whole cluster in model `Cold th45' over a timespan of several initial half-mass relaxation times, $\trhi$.  Middle and right panels: time evolution of the $\phi$ and $\theta$ components of global $\omegavec$.}
    \label{fig:omglob}
\end{figure*}

\begin{figure*}
    \includegraphics{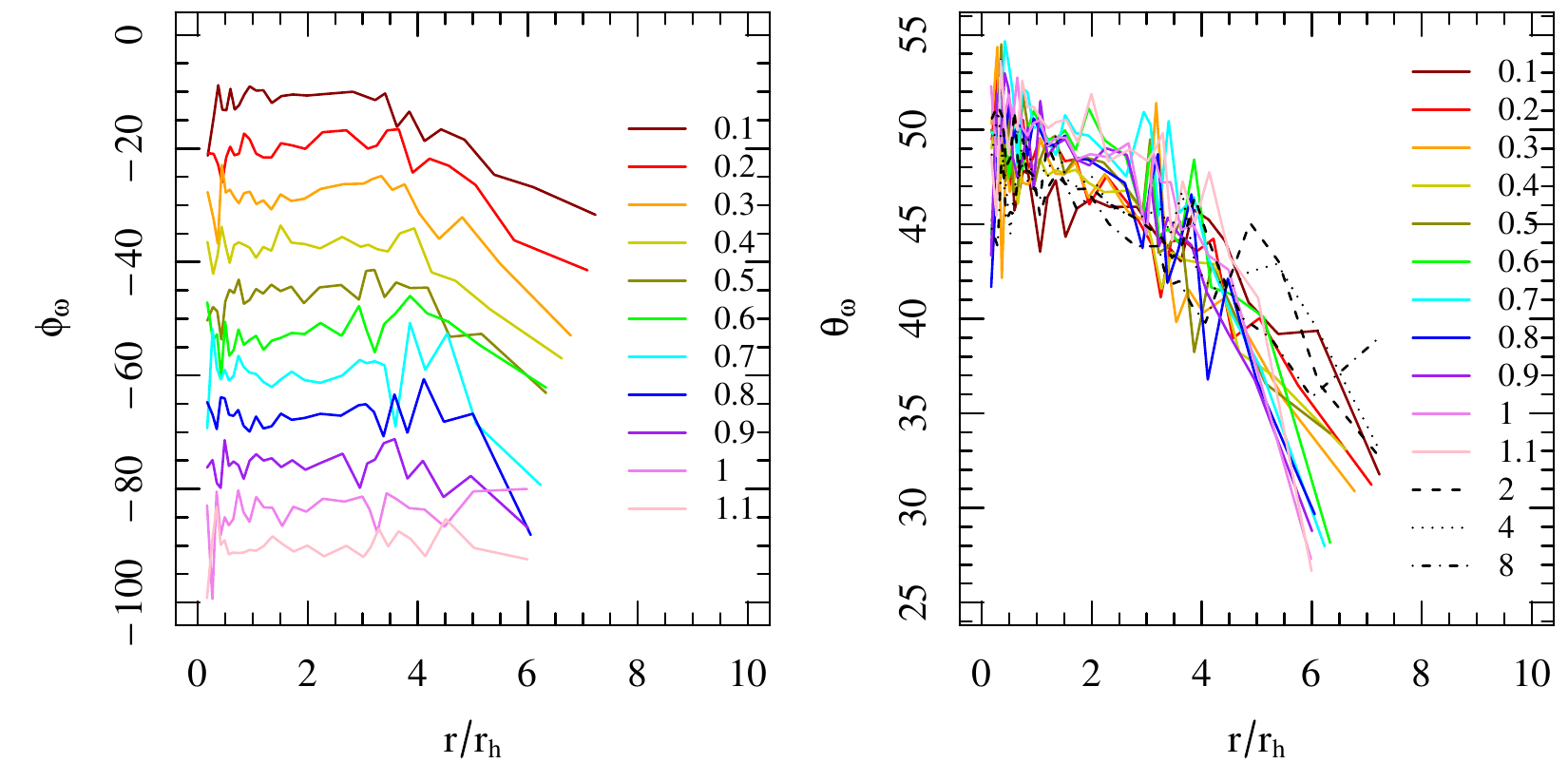}
    \caption{Left panel: time evolution of the radial profiles of $\phiomega$ over a timespan of about 1$\trhi$ (the legend of each panel is time normalized to $\trhi$).  Right panel: time evolution of the radial profiles of $\thetaomega$ over a timespan of about 1$\trhi$, but also includes profiles at 2, 4 and 8$\trhi$ to show that the radial variation remains over several relaxation times.}
    \label{fig:phithetatrh}
\end{figure*}

We begin with an overview and visualization of the violent relaxation phase of our systems.  Our systems are all initially cold homogeneous spheres of stars spinning in solid-body rotation.  They then collapse from lack of pressure support, reaching a maximum density, shortly after 1$\tff$.  The clusters then expand and eventually settle into an equilibrium configuration.  This process causes many stars to escape the cluster, and leaves a configuration characterized by a inner dense core and sparse outer halo; many of the halo stars are energetically unbound but still within the Jacobi radius of the cluster.  By conservation of angular momentum, when the cluster collapses, it rotates faster; when the cluster expands, the rotation speed of the stars that move to the outer regions slows down, creating a differential rotation within the cluster (see next section and also \citealt{vesperini2014,tiongco2017,livernois2021}).  Each clusters' final shape (specifically near the core to around the half-mass radius, $\rh$, excluding the sparse halo) at the end of the violent relaxation process is triaxial \citep[see also][]{labini2015}; the elongation of the major axis is primarily influenced by the initial amount of rotation, with faster rotating initial conditions producing more elongated systems.

For the set of models starting with $\thetaomegai$ equal to 45 degrees, Fig. \ref{fig:images1} shows the kinematics and structure of the clusters after they have  settled into an equilibrium configuration at the end of the simulation. Specifically, this figure shows three projected number density images along the $x$-, $y$-, and $z$-axes, and also three images of the line-of-sight (hereafter, los) velocities in those same projections with isodensity contours overlaid.

Fig. \ref{fig:images4} shows these same panels for simulation Cold th90, but zoomed out to emphasize the kinematical differences between the inner and the outer regions; these differences appear in all of the simulations\footnote{Animations showing the evolution of these projected number density and line-of-sight velocity images (zoomed in and out) are  available online at \url{https://mtiongco.github.io/vrrot-anims}}.

The images selected here for illustration, along with the animations, offer a representation of the complexity that emerges from our generic initial conditions, particularly if a cluster is `observed' along an arbitrary direction of the line-of-sight and evolutionary stage, and we do not assume knowledge of the initial orientation of the rotation axis. Especially in the cases where clusters start with their $\thetaomegai$ equal to 45 or 135 degrees, the cluster's triaxiality and the rotation axis not being aligned with any of the $x$, $y$, or $z$ directions, can lead to projected configurations characterised by a variety of aspect ratios and apparent axes about which the cluster rotates.  The simulations also show that the los velocities are not all aligned along one rotation axis, with a larger deviation from the initial rotation axis direction in the outer regions, where the effects of the tidal field are stronger.  Though this may not be obvious in all of the los velocity maps, we will further illustrate this point using another diagnostic in the next section.

\subsection{Three-dimensional Properties}

The complexity of the kinematics that develops during the violent relaxation phase in our clusters is reflected in the evolution of the radial profiles of $\phiomega$ and $\thetaomega$, as shown in Figs \ref{fig:phiom_pro} and \ref{fig:thetaom_pro}.  In Fig. \ref{fig:omegar}, we also quantify the amount of rotation in each system at different times.  In all these figures, at each time the cluster is divided into spherical shells in which the angular velocity vector, $\omegavec$ is calculated by solving the equation $\mathbfit{L} = \mathbfit{I} \omegavec$, where $\mathbfit{L}$ is the angular momentum vector and $\mathbfit{I}$ is the moment of inertia tensor.  In Fig. \ref{fig:omegar}, we illustrate, as a function of the spherical radius, the rotational velocity which results from the product of the magnitude of the angular velocity vector by the radius of the spherical shell. These profiles are intended to represent the ordered motions averaged under the assumption of spherical symmetry.

These figures show that each system has differential rotation (Fig. \ref{fig:omegar}), but between each spherical shell the direction of the rotation axis varies in \textit{both} $\phiomega$ and $\thetaomega$, and these radial variations also change throughout the simulation.  The radial variation in $\phiomega$ distinguishes these results from the simulations in \citet[][which did not include an analysis of the violent relaxation phase]{tiongco2018}, where no radial variation in $\phiomega$ was observed.  Such radial variation in $\phiomega$ introduces a twist of the alignment of the rotation axis throughout the spatial extension of the system.  It can also be observed in these figures that the evolution in the radial variations is most significant in the outermost regions of the cluster, sweeping over a large range of angles, while in the inner regions, the rotation axis directions stay closer to their original values.  

At the end of the violent relaxation process, $\omegavec$ in the outermost regions of each cluster points closer to the direction of the orbital angular velocity vector than it did initially. Such a feature is more evident in clusters that begin with $\thetaomegai = 45$, while in clusters that begin with $\thetaomegai = 135$ the outermost regions (up to the Jacobi radius) tend to point into the orbital plane ($\thetaomega \approx 90$) by the end of the simulation. 

Next, we proceed in the presentation of our results with a discussion of the structure of the systems' equilibrium properties emerging at the end of the violent relaxation process.  After the system relaxes into an approximately steady state, where structural properties such as the half-mass radius, $\rh$, no longer change significantly on the free-fall timescale, we calculate the ratios of the principal semi-axes $a$, $b$, and $c$ (in order from longest to shortest) in ellipsoidal shells at different radii throughout the cluster, where the method to calculate the shape of each shell is detailed in \citet[method S1]{zemp2011}. Briefly, the principal semi-axes are found by diagonalizing the shape tensor, 

\begin{equation}
\mathbfit{S}_{ij} = \frac{\sum_{k} m_k (\mathbfit{r}_k)_i (\mathbfit{r}_k)_j}{\sum_{k} m_k},
\end{equation}

\noindent
of a shell of particles, and the square roots of the eigenvalues are proportional to the principal semi-axes.  This process is iterated until the ratios of the principal semi-axes converge to values that do not change by more than 0.1 percent between iterations. 

In Fig. \ref{fig:axialratio}, we show the axial ratios as a function of radius throughout each cluster, averaged over several profiles between 17.1 and 17.9 $\tff$.  In general, the clusters have a triaxial shape, with differences depending on the initial virial ratios. The `Cold' series of clusters have a deeper collapse, and thus faster rotation and flatter shape.  These clusters are also flatter in the center and rounder in the outer regions, where the transition region is about 1--3 $\rh$.  The `Even' series of clusters follows a similar trend but are overall less flat than the `Cold' series. Finally, the models in the the `Slow' series are those with morphology closest to spherical; this is a consequence of the fact that these models are those with the lowest initial rotational energy among those considered in this study.

Observational studies of nuclear star clusters and galaxies have shown misalignments of structural principal axes and kinematic rotation axes \citep{franx1988,derijke2004,geha2005,thomas2006,feldmeier2014,tsatsi2017,delpino2021}.  The method used to find the shape of ellipsoidal shells also gives the 3D orientation of the principal axes via the eigenvectors.  At many times throughout each simulation, we find that the orientation of the semi-minor axis, $c$, differs from the orientation of rotation axis, $\omegavec$.  Fig. \ref{fig:morphkin1} shows an example of this for one of the models.

In the final part of this section, we discuss the measurement of velocity dispersion anisotropy in the system.  Velocity anistotropy is often measured by the parameter,

\begin{equation}
\label{eq:beta}
\beta=1-\frac{\sigma_{\theta}^2+\sigma_{\phi}^2}{2\sigma_{\rm r}^2},
\end{equation}

\noindent
\citep[see, e.g.,][]{bt08}, and the radial profile of $\beta$ is typically calculated in spherical shells.  We show the results of using this method in Fig. \ref{fig:aniso_s90} at the end of simulation `Slow th90' in the top left panel.  For the adopted definition of $\beta$,  $\beta=0$ corresponds to an isotropic velocity distribution, while negative (positive) values of $\beta$ correspond to a tangentially (radially) anisotropic velocity distribution.  The profile of $\beta$ in the mid-to-outer regions  is consistent with previous numerical studies of violent relaxation of star clusters in an external tidal field \citep[e.g., see][]{boily1999,vesperini2014} in that  radial anisotropy has a maximum in between 2--4$\rh$, and then, going further outwards, the radial anisotropy decreases and the velocity distribution becomes almost isotropic.  The shape of the profile in the mid-to-outer regions can be explained by the expansion of the cluster during the rebounce phase of violent relaxation which creates radial orbits in the outer regions of the cluster, while presence of the tidal field makes the outermost regions of the cluster more isotropic due to the preferential escape of stars on radial orbits \citep[see, e.g.,][]{vesperini2014,tiongco2016}.

In our models, we find that the inner regions are characterized by tangential anisotropy \citep[for similar considerations in isolated models, see also][]{trenti2005}. However, this may be in part due to the inappropriate geometry associated with the choice of spherical binning for rotating systems: a spherical bin would include stars that have most of their velocity along the azimuthal direction, but also stars that do not, creating an artificially higher tangential velocity dispersion. We attempt to reconcile this by calculating $\beta$ in cylindrical shells aligning them along a few different directions.  This particular model shown in Fig. \ref{fig:aniso_s90}, `Slow th90,'  was chosen because its initial $\omegavec$ points along the positive $x$-axis.  Since the most ideal subtraction of the mean velocity occurs when the cylindrical bins are aligned with the rotation axis of the cluster, the top right panel of Fig. \ref{fig:aniso_s90} shows that using a los along the $x$-axis in the velocity anisotropy calculation has a reduced inner tangential anisotropy compared to setting the los along the $y$- or the $z$- axes.  In the last two panels of the second row of Fig. \ref{fig:aniso_s90}, we use lines of sight parallel to $\omegavec$ of stars within 1.5$\rh$ and outside of 1.5$\rh$ (the dividing point was determined from slightly further out from where the rotational radial profile approximately peaks). We find that using this binning and los reduces the tangential velocity anisotropy in $\beta$ in the innermost regions, although finding the most effective alignment of the cylindrical bin is complicated by the radial variation of the alignment of the rotation axis.  

We also show a similar analysis in a second set of panels in Fig. \ref{fig:aniso_s90} for simulation `Slow th135' where the initial $\omegavec$ is not aligned along any of the $x$-, $y$-, or $z$-axes.   The largest reduction of the tangential anisotropy in $\beta$ is obtained by adopting a los aligned with the rotation axis of the cluster.

\subsection{Projected Properties}

In order to connect our results to observational studies, we now present the results of our analysis of the cluster kinematic properties determined using projected quantities.
We first focus our attention on the velocity dispersion anisotropy: Fig. \ref{fig:aniso_s90} shows (dashed lines) the projected anisotropy parameter, $1-\sigma_{\rm T}/\sigma_{\rm R}$; $\sigma_{\rm T}$ and $\sigma_{\rm R}$ are, respectively, the tangential and radial velocity dispersion measured on the projection plane perpendiclar to the line of sight.  This parameter is measured in the same cylindrical bins used for $\beta$ in the panels after the top left one, and is effectively a different measure since it disregards the contribution of line-of-sight velocities (i.e., this measure is similar to that obtained with the velocities from proper motions).  We find that using the projected anisotropy parameter further reduces the innermost tangential anisotropy bias, eliminating it in the cases where the los is aligned with the rotation axis of the cluster.

Next, in Fig. \ref{fig:aniall}, we show all of the models together plotting the evolution of the projected anisotropy parameter, $1-\sigma_{\rm T}/\sigma_{\rm R}$.  In Fig. \ref{fig:aniall}, the line of sight is aligned with the rotation axis of the inner regions (corresponding to `los:r<1.5rh' in Fig. \ref{fig:aniso_s90}).  The anisotropy profile shows that the system is more radially anisotropic, particularly in the outer regions during the violent relaxation process, and this reflects the expansion of the cluster after the initial collapse phase.  As more stars escape from the cluster when the system is settling, the system loses some radial anisotropy, but the evolution slows down at later times in the simulation.

Finally, in Fig. \ref{fig:rotall}, we plot the evolution of the projected rotation curves for all models from the same line of sight, chosen to be the $z$-axis.  These radial profiles are calculated from tangential velocities in the plane perpendicular to the chosen line of sight (equivalent to rotation curves calculated from proper motions in observations), making this figure different from Fig. \ref{fig:omegar} where $\omega r$ was calculated from 3d velocities, and the rotation axis direction was not constant with radius.   Fig. \ref{fig:rotall} shows that a variety of rotation curve shapes can be obtained looking at the cluster along different lines of sight: the `th45' and the 'th135' series differ in the direction of rotation  but both show a typical rotation curve characterized by rotational velocities increasing in the inner regions and decreasing in the outer regions; the `th90' series, on the other hand, shows very small velocities due to line of sight being almost completely misaligned  from the rotation axis.  These results highlight the importance of obtaining the full phase space information to accurately measure, for example, the angular momentum in a system.

\subsection{Long-term Evolution}

In this section, we present the results of a simulation following the long-term evolution driven by two-body relaxation of our model, `Cold th45'. The goal here is to explore how the various complexities emerging at the end of the violent relaxation phase are affected by the cluster's long-term evolution and compare these results with those found in \citet{tiongco2018}.

One of the major results of \citet{tiongco2018} was that the rotation axis of the cluster precesses due to the torque of the host galaxy. In Fig. \ref{fig:omglob}, we show the evolution of the Cartesian components of the global $\omegavec$ for `Cold th45', where the precession about the $z$-axis is indicated by the oscillation of $x$- and $y$- components, and along with the almost linear change with time in $\phiomega$. Also in line with our previous results is the gradual loss of angular velocity (rotation) over several relaxation times, as indicated by the magnitude of  $\omega_{\rm z}$ and amplitude of the oscillations in $\omega_{\rm x}$ and $\omega_{\rm y}$ decreasing.  The third panel of Fig. \ref{fig:omglob} shows the evolution of $\thetaomega$ decreasing a few degrees due to the loss of intrinsic angular momentum and the system moving towards a tidally locked configuration in which the rotation of the cluster would be predominantly about the $z$-axis.

We also examine the long-term evolution of the profiles $\phiomega$ and $\thetaomega$; as discussed in the previous sections, at the end of the violent relaxation phase the cluster is characterized by a radial variation in $\phiomega$ and $\thetaomega$.   The left panel of Fig. \ref{fig:phithetatrh} shows that the radial variation in $\phiomega$ is gradually erased over the two-body relaxation timescale (the profiles for $\phiomega$ move down along the plot due to the precession of the cluster's rotation axis as mentioned previously).  On the other hand, the other panel shows the evolution of $\thetaomega$, where the radial variation remains over several two-body relaxation timescales.

\section{Conclusions}

We have explored via \Nbody simulations the early dynamical evolution of rotating star clusters undergoing violent relaxation under the effects of the tidal field of their host galaxy.  By allowing the clusters to initially have a rotation axis orientation not aligned with their orbital angular velocity vector, we have found that the equilibrium configurations reached at the end of the violent relaxation phase are characterized by a variety of complex structural and kinematic features.  

We have explored nine different initial conditions varying in initial virial ratio in random and rotational kinetic energy: the `Even' series having the same virial ratio in random and rotational kinetic energy, the `Cold' series having a lower random virial ratio, and the `Slow' series having a lower rotational virial ratio.  We have also explored the evolution of systems with different values  of  the initial angle between the cluster's rotation axis  and the cluster's orbital angular velocity vector (45 degrees, 90 degrees, and 135 degrees).

The violent relaxation process leaves the cluster with a differential rotation curve characterized by a rotational velocity rising in the inner regions, peaking in the intermediate regions, and then falling in the outer parts of the cluster.  

Most notably, we find that the clusters end up with a radial variation in rotation axis direction along both the polar angle $\theta$ and the azimuthal angle $\phi$ directions.  During the violent relaxation process, the rotation axis directions oscillate along the $\theta$ and $\phi$ directions, with a stronger amplitude in the outermost regions.  The radial variation in the rotation axis directions is found also in the final equilibrium configuration at the end of the violent relaxation process, with the outermost regions pointing closer to the direction of the orbital angular velocity vector.

Our clusters develop a triaxial morphology with the models in the `Cold' series being those showing the strongest flattening; this is a consequence of the fact that the models of this series are those that undergo the deepest collapse and emerge from the violent relaxation with the most rapid rotation.  We find a radial variation in the morphological minor axis position angle, but this position angle does not necessarily align with the rotation axis.

We also explored how the complex features of these clusters appear in projection as they do in observational studies.  From different perspectives, a radial velocity field image combined with photometric isophotes may look like the cluster is rotating about its major, minor, or another axis altogether.  The measurement of velocity dispersion anisotropy may have a tangential bias near the centre of the cluster due to a suboptimal orientation of the line of sight (i.e., a los that is not close to the rotation axis of cluster) that can lead to anisotropy measurements contaminated by ordered rotational motion.  The tangential bias may be reduced when the projected anisotropy parameter is measured on the plane normal to the mean position of the rotation axis; however, the tangential bias may not be completely eliminated due to the rotation axis positional variation throughout the cluster.

Finally, we have followed the evolution of one of our clusters over a timescale of several half-mass relaxation times, and found that the radial variation of the rotation axis position along the azimuthal  direction is erased  over a relaxation time.  Thus, observing a radial variation of the position angle of the rotation axis along the azimuthal direction of a star cluster may be a kinematical signature of a dynamically young cluster.  On the other hand, the radial variation of the rotation axis position along the polar $\theta$ direction is not erased during the cluster's long-term evolution. This is a consequence of the evolution towards a tidally locked rotation that drives  the evolution of the cluster towards rotation about the $z$-axis (i.e. parallel to the orbital angular velocity vector of the cluster).

Current morphological and kinematical profiles of Galactic globular clusters do not yet have the ability to fully identify the subtle radial variations we have highlighted here, mostly due to the astrometric limitations in the reconstruction of the 3D velocity space of objects characterised by such crowded fields. Yet, we wish to note that a detailed study of the Sagittarius dwarf galaxy by \citet{delpino2021} has recently brought to light the first evidence of precession and nutation motions imparted by the Milky Way tidal field onto a satellite stellar system. We are therefore hopeful that forthcoming Gaia data releases will enable this kind of refined investigations also for local globular clusters. More generally, the complex structural and kinematic features revealed by our study  provide new elements for the interpretation of observations of star clusters and can shed light on their formation and dynamical history.  In future studies, we will extend the investigation presented in this paper to consider more realistic galactic potentials and the effects of a time-dependent external potential for clusters on non-circular orbits.

\section*{Acknowledgements}

MT thanks the Eccentric Dynamics research group at CU Boulder for feedback on the results and presentation.  This research was supported in part by Lilly Endowment, Inc., through its support for the Indiana University Pervasive Technology Institute, and in part by the Indiana METACyt Initiative. The Indiana METACyt Initiative at IU is also supported in part by Lilly Endowment, Inc.  ALV acknowledges support from a UKRI Future Leaders Fellowship (MR/S018859/1).

\section*{Data Availability}

The data underlying this article will be shared on reasonable request to the corresponding author.




\bibliographystyle{mnras}
\bibliography{main} 

\begin{thebibliography}{}
\makeatletter
\relax
\def\mn@urlcharsother{\let\do\@makeother \do\$\do\&\do\#\do\^\do\_\do\%\do\~}
\def\mn@doi{\begingroup\mn@urlcharsother \@ifnextchar [ {\mn@doi@}
  {\mn@doi@[]}}
\def\mn@doi@[#1]#2{\def\@tempa{#1}\ifx\@tempa\@empty \href
  {http://dx.doi.org/#2} {doi:#2}\else \href {http://dx.doi.org/#2} {#1}\fi
  \endgroup}
\def\mn@eprint#1#2{\mn@eprint@#1:#2::\@nil}
\def\mn@eprint@arXiv#1{\href {http://arxiv.org/abs/#1} {{\tt arXiv:#1}}}
\def\mn@eprint@dblp#1{\href {http://dblp.uni-trier.de/rec/bibtex/#1.xml}
  {dblp:#1}}
\def\mn@eprint@#1:#2:#3:#4\@nil{\def\@tempa {#1}\def\@tempb {#2}\def\@tempc
  {#3}\ifx \@tempc \@empty \let \@tempc \@tempb \let \@tempb \@tempa \fi \ifx
  \@tempb \@empty \def\@tempb {arXiv}\fi \@ifundefined
  {mn@eprint@\@tempb}{\@tempb:\@tempc}{\expandafter \expandafter \csname
  mn@eprint@\@tempb\endcsname \expandafter{\@tempc}}}

\bibitem[\protect\citeauthoryear{{Aarseth}}{{Aarseth}}{2003}]{aarseth2003}
{Aarseth} S.~J.,  2003, {Gravitational N-Body Simulations}.
Cambridge University Press

\bibitem[\protect\citeauthoryear{{Aarseth}, {Lin}  \& {Papaloizou}}{{Aarseth}
  et~al.}{1988}]{aarseth1988}
{Aarseth} S.~J.,  {Lin} D.~N.~C.,   {Papaloizou} J.~C.~B.,  1988, \mn@doi
  [\apj] {10.1086/165895}, \href
  {https://ui.adsabs.harvard.edu/abs/1988ApJ...324..288A} {324, 288}

\bibitem[\protect\citeauthoryear{{Aguilar} \& {Merritt}}{{Aguilar} \&
  {Merritt}}{1990}]{aguilar1990}
{Aguilar} L.~A.,  {Merritt} D.,  1990, \mn@doi [\apj] {10.1086/168665}, \href
  {https://ui.adsabs.harvard.edu/abs/1990ApJ...354...33A} {354, 33}

\bibitem[\protect\citeauthoryear{{Armstrong}, {Wright}, {Jeffries}  \&
  {Jackson}}{{Armstrong} et~al.}{2020}]{armstrong2020}
{Armstrong} J.~J.,  {Wright} N.~J.,  {Jeffries} R.~D.,   {Jackson} R.~J.,
  2020, \mn@doi [\mnras] {10.1093/mnras/staa939}, \href
  {https://ui.adsabs.harvard.edu/abs/2020MNRAS.494.4794A} {494, 4794}

\bibitem[\protect\citeauthoryear{{Ballone}, {Torniamenti}, {Mapelli}, {Di
  Carlo}, {Spera}, {Rastello}, {Gaspari}  \& {Iorio}}{{Ballone}
  et~al.}{2021}]{ballone2021}
{Ballone} A.,  {Torniamenti} S.,  {Mapelli} M.,  {Di Carlo} U.~N.,  {Spera} M.,
   {Rastello} S.,  {Gaspari} N.,   {Iorio} G.,  2021, \mn@doi [\mnras]
  {10.1093/mnras/staa3763}, \href
  {https://ui.adsabs.harvard.edu/abs/2021MNRAS.501.2920B} {501, 2920}

\bibitem[\protect\citeauthoryear{{Bellini}, {Bianchini}, {Varri}, {Anderson},
  {Piotto}, {van der Marel}, {Vesperini}  \& {Watkins}}{{Bellini}
  et~al.}{2017}]{bellini2017}
{Bellini} A.,  {Bianchini} P.,  {Varri} A.~L.,  {Anderson} J.,  {Piotto} G.,
  {van der Marel} R.~P.,  {Vesperini} E.,   {Watkins} L.~L.,  2017, \mn@doi
  [\apj] {10.3847/1538-4357/aa7c5f}, \href
  {https://ui.adsabs.harvard.edu/abs/2017ApJ...844..167B} {844, 167}

\bibitem[\protect\citeauthoryear{{Bianchini}, {Varri}, {Bertin}  \&
  {Zocchi}}{{Bianchini} et~al.}{2013}]{bianchini2013}
{Bianchini} P.,  {Varri} A.~L.,  {Bertin} G.,   {Zocchi} A.,  2013, \mn@doi
  [\apj] {10.1088/0004-637X/772/1/67}, \href
  {http://adsabs.harvard.edu/abs/2013ApJ...772...67B} {772, 67}

\bibitem[\protect\citeauthoryear{{Bianchini}, {van der Marel}, {del Pino},
  {Watkins}, {Bellini}, {Fardal}, {Libralato}  \& {Sills}}{{Bianchini}
  et~al.}{2018}]{bianchini2018}
{Bianchini} P.,  {van der Marel} R.~P.,  {del Pino} A.,  {Watkins} L.~L.,
  {Bellini} A.,  {Fardal} M.~A.,  {Libralato} M.,   {Sills} A.,  2018, \mn@doi
  [\mnras] {10.1093/mnras/sty2365}, \href
  {https://ui.adsabs.harvard.edu/abs/2018MNRAS.481.2125B} {481, 2125}

\bibitem[\protect\citeauthoryear{{Binney} \& {Tremaine}}{{Binney} \&
  {Tremaine}}{2008}]{bt08}
{Binney} J.,  {Tremaine} S.,  2008, {Galactic Dynamics: Second Edition}.
Princeton University Press

\bibitem[\protect\citeauthoryear{{Boberg}, {Vesperini}, {Friel}, {Tiongco}  \&
  {Varri}}{{Boberg} et~al.}{2017}]{boberg2017}
{Boberg} O.~M.,  {Vesperini} E.,  {Friel} E.~D.,  {Tiongco} M.~A.,   {Varri}
  A.~L.,  2017, \mn@doi [\apj] {10.3847/1538-4357/aa7070}, \href
  {http://adsabs.harvard.edu/abs/2017ApJ...841..114B} {841, 114}

\bibitem[\protect\citeauthoryear{{Boily}, {Clarke}  \& {Murray}}{{Boily}
  et~al.}{1999}]{boily1999}
{Boily} C.~M.,  {Clarke} C.~J.,   {Murray} S.~D.,  1999, \mn@doi [\mnras]
  {10.1046/j.1365-8711.1999.02153.x}, \href
  {https://ui.adsabs.harvard.edu/abs/1999MNRAS.302..399B} {302, 399}

\bibitem[\protect\citeauthoryear{{Chen}, {Li}  \& {Vogelsberger}}{{Chen}
  et~al.}{2021}]{chen2021}
{Chen} Y.,  {Li} H.,   {Vogelsberger} M.,  2021, \mn@doi [\mnras]
  {10.1093/mnras/stab491}, \href
  {https://ui.adsabs.harvard.edu/abs/2021MNRAS.502.6157C} {502, 6157}

\bibitem[\protect\citeauthoryear{{Dalessandro}, {Raso}, {Kamann}, {Bellazzini},
  {Vesperini}, {Bellini}  \& {Beccari}}{{Dalessandro}
  et~al.}{2021a}]{dalessandro2021b}
{Dalessandro} E.,  {Raso} S.,  {Kamann} S.,  {Bellazzini} M.,  {Vesperini} E.,
  {Bellini} A.,   {Beccari} G.,  2021a, \mn@doi [\mnras]
  {10.1093/mnras/stab1257}, \href
  {https://ui.adsabs.harvard.edu/abs/2021MNRAS.506..813D} {506, 813}

\bibitem[\protect\citeauthoryear{{Dalessandro} et~al.,}{{Dalessandro}
  et~al.}{2021b}]{dalessandro2021a}
{Dalessandro} E.,  et~al., 2021b, \mn@doi [\apj] {10.3847/1538-4357/abda43},
  \href {https://ui.adsabs.harvard.edu/abs/2021ApJ...909...90D} {909, 90}

\bibitem[\protect\citeauthoryear{{De Rijcke}, {Dejonghe}, {Zeilinger}  \&
  {Hau}}{{De Rijcke} et~al.}{2004}]{derijke2004}
{De Rijcke} S.,  {Dejonghe} H.,  {Zeilinger} W.~W.,   {Hau} G.~K.~T.,  2004,
  \mn@doi [\aap] {10.1051/0004-6361:20041205}, \href
  {https://ui.adsabs.harvard.edu/abs/2004A&A...426...53D} {426, 53}

\bibitem[\protect\citeauthoryear{{Einsel} \& {Spurzem}}{{Einsel} \&
  {Spurzem}}{1999}]{einsel1999}
{Einsel} C.,  {Spurzem} R.,  1999, \mn@doi [\mnras]
  {10.1046/j.1365-8711.1999.02083.x}, \href
  {http://adsabs.harvard.edu/abs/1999MNRAS.302...81E} {302, 81}

\bibitem[\protect\citeauthoryear{{Ernst}, {Glaschke}, {Fiestas}, {Just}  \&
  {Spurzem}}{{Ernst} et~al.}{2007}]{ernst2007}
{Ernst} A.,  {Glaschke} P.,  {Fiestas} J.,  {Just} A.,   {Spurzem} R.,  2007,
  \mn@doi [\mnras] {10.1111/j.1365-2966.2007.11602.x}, \href
  {http://adsabs.harvard.edu/abs/2007MNRAS.377..465E} {377, 465}

\bibitem[\protect\citeauthoryear{{Fabricius} et~al.,}{{Fabricius}
  et~al.}{2014}]{fabricius2014}
{Fabricius} M.~H.,  et~al., 2014, \mn@doi [\apjl]
  {10.1088/2041-8205/787/2/L26}, \href
  {https://ui.adsabs.harvard.edu/abs/2014ApJ...787L..26F} {787, L26}

\bibitem[\protect\citeauthoryear{{Feldmeier} et~al.,}{{Feldmeier}
  et~al.}{2014}]{feldmeier2014}
{Feldmeier} A.,  et~al., 2014, \mn@doi [\aap] {10.1051/0004-6361/201423777},
  \href {https://ui.adsabs.harvard.edu/abs/2014A&A...570A...2F} {570, A2}

\bibitem[\protect\citeauthoryear{{Ferraro} et~al.,}{{Ferraro}
  et~al.}{2018}]{Ferraro2018}
{Ferraro} F.~R.,  et~al., 2018, \mn@doi [\apj] {10.3847/1538-4357/aabe2f},
  \href {https://ui.adsabs.harvard.edu/abs/2018ApJ...860...50F} {860, 50}

\bibitem[\protect\citeauthoryear{{Franx} \& {Illingworth}}{{Franx} \&
  {Illingworth}}{1988}]{franx1988}
{Franx} M.,  {Illingworth} G.~D.,  1988, \mn@doi [\apjl] {10.1086/185139},
  \href {https://ui.adsabs.harvard.edu/abs/1988ApJ...327L..55F} {327, L55}

\bibitem[\protect\citeauthoryear{{Gebhardt}, {Pryor}, {O'Connell}, {Williams}
  \& {Hesser}}{{Gebhardt} et~al.}{2000}]{gebhardt2000}
{Gebhardt} K.,  {Pryor} C.,  {O'Connell} R.~D.,  {Williams} T.~B.,   {Hesser}
  J.~E.,  2000, \mn@doi [\aj] {10.1086/301275}, \href
  {http://adsabs.harvard.edu/abs/2000AJ....119.1268G} {119, 1268}

\bibitem[\protect\citeauthoryear{{Geha}, {Guhathakurta}  \& {van der
  Marel}}{{Geha} et~al.}{2005}]{geha2005}
{Geha} M.,  {Guhathakurta} P.,   {van der Marel} R.~P.,  2005, \mn@doi [\aj]
  {10.1086/430188}, \href
  {https://ui.adsabs.harvard.edu/abs/2005AJ....129.2617G} {129, 2617}

\bibitem[\protect\citeauthoryear{{Gott}}{{Gott}}{1973}]{gott1973}
{Gott} Richard~J. I.,  1973, \mn@doi [\apj] {10.1086/152514}, \href
  {https://ui.adsabs.harvard.edu/abs/1973ApJ...186..481G} {186, 481}

\bibitem[\protect\citeauthoryear{{Heggie} \& {Hut}}{{Heggie} \&
  {Hut}}{2003}]{heggie2003}
{Heggie} D.,  {Hut} P.,  2003, {The Gravitational Million-Body Problem}.
Cambridge University Press

\bibitem[\protect\citeauthoryear{{Heisler} \& {Tremaine}}{{Heisler} \&
  {Tremaine}}{1986}]{heisler1986}
{Heisler} J.,  {Tremaine} S.,  1986, \mn@doi [\icarus]
  {10.1016/0019-1035(86)90060-6}, \href
  {https://ui.adsabs.harvard.edu/abs/1986Icar...65...13H} {65, 13}

\bibitem[\protect\citeauthoryear{{Hohl} \& {Zang}}{{Hohl} \&
  {Zang}}{1979}]{hohl1979}
{Hohl} F.,  {Zang} T.~A.,  1979, \mn@doi [\aj] {10.1086/112454}, \href
  {https://ui.adsabs.harvard.edu/abs/1979AJ.....84..585H} {84, 585}

\bibitem[\protect\citeauthoryear{{Hong}, {Kim}, {Lee}  \& {Spurzem}}{{Hong}
  et~al.}{2013}]{hong2013}
{Hong} J.,  {Kim} E.,  {Lee} H.~M.,   {Spurzem} R.,  2013, \mn@doi [\mnras]
  {10.1093/mnras/stt099}, \href
  {http://adsabs.harvard.edu/abs/2013MNRAS.430.2960H} {430, 2960}

\bibitem[\protect\citeauthoryear{{Jerabkova}, {Boffin}, {Beccari}, {de Marchi},
  {de Bruijne}  \& {Prusti}}{{Jerabkova} et~al.}{2021}]{jerabkova2021}
{Jerabkova} T.,  {Boffin} H. M.~J.,  {Beccari} G.,  {de Marchi} G.,  {de
  Bruijne} J. H.~J.,   {Prusti} T.,  2021, \mn@doi [\aap]
  {10.1051/0004-6361/202039949}, \href
  {https://ui.adsabs.harvard.edu/abs/2021A&A...647A.137J} {647, A137}

\bibitem[\protect\citeauthoryear{{Kamann} et~al.,}{{Kamann}
  et~al.}{2018}]{kamann2018}
{Kamann} S.,  et~al., 2018, \mn@doi [\mnras] {10.1093/mnras/stx2719}, \href
  {https://ui.adsabs.harvard.edu/abs/2018MNRAS.473.5591K} {473, 5591}

\bibitem[\protect\citeauthoryear{{Kim}, {Einsel}, {Lee}, {Spurzem}  \&
  {Lee}}{{Kim} et~al.}{2002}]{kim2002}
{Kim} E.,  {Einsel} C.,  {Lee} H.~M.,  {Spurzem} R.,   {Lee} M.~G.,  2002,
  \mn@doi [\mnras] {10.1046/j.1365-8711.2002.05420.x}, \href
  {http://adsabs.harvard.edu/abs/2002MNRAS.334..310K} {334, 310}

\bibitem[\protect\citeauthoryear{{Kim}, {Lee}  \& {Spurzem}}{{Kim}
  et~al.}{2004}]{kim2004}
{Kim} E.,  {Lee} H.~M.,   {Spurzem} R.,  2004, \mn@doi [\mnras]
  {10.1111/j.1365-2966.2004.07776.x}, \href
  {http://adsabs.harvard.edu/abs/2004MNRAS.351..220K} {351, 220}

\bibitem[\protect\citeauthoryear{{Kim}, {Yoon}, {Lee}  \& {Spurzem}}{{Kim}
  et~al.}{2008}]{kim2008}
{Kim} E.,  {Yoon} I.,  {Lee} H.~M.,   {Spurzem} R.,  2008, \mn@doi [\mnras]
  {10.1111/j.1365-2966.2007.12524.x}, \href
  {http://adsabs.harvard.edu/abs/2008MNRAS.383....2K} {383, 2}

\bibitem[\protect\citeauthoryear{{Kuhn}, {Hillenbrand}, {Sills}, {Feigelson}
  \& {Getman}}{{Kuhn} et~al.}{2019}]{kuhn2019}
{Kuhn} M.~A.,  {Hillenbrand} L.~A.,  {Sills} A.,  {Feigelson} E.~D.,   {Getman}
  K.~V.,  2019, \mn@doi [\apj] {10.3847/1538-4357/aaef8c}, \href
  {https://ui.adsabs.harvard.edu/abs/2019ApJ...870...32K} {870, 32}

\bibitem[\protect\citeauthoryear{{Lah{\'e}n}, {Naab}, {Johansson}, {Elmegreen},
  {Hu}  \& {Walch}}{{Lah{\'e}n} et~al.}{2020}]{lahen2020}
{Lah{\'e}n} N.,  {Naab} T.,  {Johansson} P.~H.,  {Elmegreen} B.,  {Hu} C.-Y.,
  {Walch} S.,  2020, \mn@doi [\apj] {10.3847/1538-4357/abc001}, \href
  {https://ui.adsabs.harvard.edu/abs/2020ApJ...904...71L} {904, 71}

\bibitem[\protect\citeauthoryear{{Lanzoni} et~al.,}{{Lanzoni}
  et~al.}{2018a}]{Lanzoni2018}
{Lanzoni} B.,  et~al., 2018a, \mn@doi [\apj] {10.3847/1538-4357/aac26a}, \href
  {https://ui.adsabs.harvard.edu/abs/2018ApJ...861...16L} {861, 16}

\bibitem[\protect\citeauthoryear{{Lanzoni} et~al.,}{{Lanzoni}
  et~al.}{2018b}]{Lanzoni2018b}
{Lanzoni} B.,  et~al., 2018b, \mn@doi [\apj] {10.3847/1538-4357/aad810}, \href
  {https://ui.adsabs.harvard.edu/abs/2018ApJ...865...11L} {865, 11}

\bibitem[\protect\citeauthoryear{{Lim}, {Hong}, {Yun}, {Hwang}, {Kim}, {Lee},
  {Park}  \& {Park}}{{Lim} et~al.}{2020}]{lim2020}
{Lim} B.,  {Hong} J.,  {Yun} H.-S.,  {Hwang} N.,  {Kim} J.~S.,  {Lee} J.-E.,
  {Park} B.-G.,   {Park} S.,  2020, \mn@doi [\apj] {10.3847/1538-4357/aba0a3},
  \href {https://ui.adsabs.harvard.edu/abs/2020ApJ...899..121L} {899, 121}

\bibitem[\protect\citeauthoryear{{Livernois}, {Vesperini}, {Tiongco}, {Varri}
  \& {Dalessandro}}{{Livernois} et~al.}{2021}]{livernois2021}
{Livernois} A.,  {Vesperini} E.,  {Tiongco} M.,  {Varri} A.~L.,   {Dalessandro}
  E.,  2021, \mn@doi [\mnras] {10.1093/mnras/stab2119}, \href
  {https://ui.adsabs.harvard.edu/abs/2021MNRAS.tmp.1924L} {}

\bibitem[\protect\citeauthoryear{{Lynden-Bell}}{{Lynden-Bell}}{1967}]{lyndenbell1967}
{Lynden-Bell} D.,  1967, \mn@doi [\mnras] {10.1093/mnras/136.1.101}, \href
  {https://ui.adsabs.harvard.edu/abs/1967MNRAS.136..101L} {136, 101}

\bibitem[\protect\citeauthoryear{{Mapelli}}{{Mapelli}}{2017}]{mapelli2017}
{Mapelli} M.,  2017, \mn@doi [\mnras] {10.1093/mnras/stx304}, \href
  {https://ui.adsabs.harvard.edu/abs/2017MNRAS.467.3255M} {467, 3255}

\bibitem[\protect\citeauthoryear{{Nitadori} \& {Aarseth}}{{Nitadori} \&
  {Aarseth}}{2012}]{nitadori2012}
{Nitadori} K.,  {Aarseth} S.~J.,  2012, \mn@doi [\mnras]
  {10.1111/j.1365-2966.2012.21227.x}, \href
  {https://ui.adsabs.harvard.edu/abs/2012MNRAS.424..545N} {424, 545}

\bibitem[\protect\citeauthoryear{{Sollima}, {Baumgardt}  \& {Hilker}}{{Sollima}
  et~al.}{2019}]{sollima2019}
{Sollima} A.,  {Baumgardt} H.,   {Hilker} M.,  2019, \mn@doi [\mnras]
  {10.1093/mnras/stz505}, \href
  {https://ui.adsabs.harvard.edu/abs/2019MNRAS.485.1460S} {485, 1460}

\bibitem[\protect\citeauthoryear{{Sylos Labini}, {Benhaiem}  \& {Joyce}}{{Sylos
  Labini} et~al.}{2015}]{labini2015}
{Sylos Labini} F.,  {Benhaiem} D.,   {Joyce} M.,  2015, \mn@doi [\mnras]
  {10.1093/mnras/stv581}, \href
  {https://ui.adsabs.harvard.edu/abs/2015MNRAS.449.4458S} {449, 4458}

\bibitem[\protect\citeauthoryear{{Thomas}, {Brimioulle}, {Bender}, {Hopp},
  {Greggio}, {Maraston}  \& {Saglia}}{{Thomas} et~al.}{2006}]{thomas2006}
{Thomas} D.,  {Brimioulle} F.,  {Bender} R.,  {Hopp} U.,  {Greggio} L.,
  {Maraston} C.,   {Saglia} R.~P.,  2006, \mn@doi [\aap]
  {10.1051/0004-6361:200500215}, \href
  {https://ui.adsabs.harvard.edu/abs/2006A&A...445L..19T} {445, L19}

\bibitem[\protect\citeauthoryear{{Tiongco}, {Vesperini}  \& {Varri}}{{Tiongco}
  et~al.}{2016}]{tiongco2016}
{Tiongco} M.~A.,  {Vesperini} E.,   {Varri} A.~L.,  2016, \mn@doi [\mnras]
  {10.1093/mnras/stv2574}, \href
  {https://ui.adsabs.harvard.edu/abs/2016MNRAS.455.3693T} {455, 3693}

\bibitem[\protect\citeauthoryear{{Tiongco}, {Vesperini}  \& {Varri}}{{Tiongco}
  et~al.}{2017}]{tiongco2017}
{Tiongco} M.~A.,  {Vesperini} E.,   {Varri} A.~L.,  2017, \mn@doi [\mnras]
  {10.1093/mnras/stx853}, \href
  {https://ui.adsabs.harvard.edu/abs/2017MNRAS.469..683T} {469, 683}

\bibitem[\protect\citeauthoryear{{Tiongco}, {Vesperini}  \& {Varri}}{{Tiongco}
  et~al.}{2018}]{tiongco2018}
{Tiongco} M.~A.,  {Vesperini} E.,   {Varri} A.~L.,  2018, \mn@doi [\mnras]
  {10.1093/mnrasl/sly009}, \href
  {https://ui.adsabs.harvard.edu/abs/2018MNRAS.475L..86T} {475, L86}

\bibitem[\protect\citeauthoryear{{Tiongco}, {Vesperini}  \& {Varri}}{{Tiongco}
  et~al.}{2019}]{tiongco2019}
{Tiongco} M.~A.,  {Vesperini} E.,   {Varri} A.~L.,  2019, \mn@doi [\mnras]
  {10.1093/mnras/stz1595}, \href
  {https://ui.adsabs.harvard.edu/abs/2019MNRAS.487.5535T} {487, 5535}

\bibitem[\protect\citeauthoryear{{Trenti}, {Bertin}  \& {van Albada}}{{Trenti}
  et~al.}{2005}]{trenti2005}
{Trenti} M.,  {Bertin} G.,   {van Albada} T.~S.,  2005, \mn@doi [\aap]
  {10.1051/0004-6361:20041705}, \href
  {https://ui.adsabs.harvard.edu/abs/2005A&A...433...57T} {433, 57}

\bibitem[\protect\citeauthoryear{{Tsatsi}, {Mastrobuono-Battisti}, {van de
  Ven}, {Perets}, {Bianchini}  \& {Neumayer}}{{Tsatsi}
  et~al.}{2017}]{tsatsi2017}
{Tsatsi} A.,  {Mastrobuono-Battisti} A.,  {van de Ven} G.,  {Perets} H.~B.,
  {Bianchini} P.,   {Neumayer} N.,  2017, \mn@doi [\mnras]
  {10.1093/mnras/stw2593}, \href
  {https://ui.adsabs.harvard.edu/abs/2017MNRAS.464.3720T} {464, 3720}

\bibitem[\protect\citeauthoryear{{Usher}, {Kamann}, {Gieles},
  {H{\'e}nault-Brunet}, {Dalessandro}, {Balbinot}  \& {Sollima}}{{Usher}
  et~al.}{2021}]{usher2021}
{Usher} C.,  {Kamann} S.,  {Gieles} M.,  {H{\'e}nault-Brunet} V.,
  {Dalessandro} E.,  {Balbinot} E.,   {Sollima} A.,  2021, \mn@doi [\mnras]
  {10.1093/mnras/stab565}, \href
  {https://ui.adsabs.harvard.edu/abs/2021MNRAS.tmp..581U} {}

\bibitem[\protect\citeauthoryear{{Vasiliev}}{{Vasiliev}}{2019}]{Vasiliev2019}
{Vasiliev} E.,  2019, \mn@doi [\mnras] {10.1093/mnras/stz2100}, \href
  {https://ui.adsabs.harvard.edu/abs/2019MNRAS.489..623V} {489, 623}

\bibitem[\protect\citeauthoryear{{Vesperini}, {Varri}, {McMillan}  \&
  {Zepf}}{{Vesperini} et~al.}{2014}]{vesperini2014}
{Vesperini} E.,  {Varri} A.~L.,  {McMillan} S.~L.~W.,   {Zepf} S.~E.,  2014,
  \mn@doi [\mnras] {10.1093/mnrasl/slu088}, \href
  {https://ui.adsabs.harvard.edu/abs/2014MNRAS.443L..79V} {443, L79}

\bibitem[\protect\citeauthoryear{{Vitral} \& {Mamon}}{{Vitral} \&
  {Mamon}}{2021}]{vitral2021}
{Vitral} E.,  {Mamon} G.~A.,  2021, \mn@doi [\aap]
  {10.1051/0004-6361/202039650}, \href
  {https://ui.adsabs.harvard.edu/abs/2021A&A...646A..63V} {646, A63}

\bibitem[\protect\citeauthoryear{{Wan} et~al.,}{{Wan} et~al.}{2021}]{wan2021}
{Wan} Z.,  et~al., 2021, \mn@doi [\mnras] {10.1093/mnras/stab306}, \href
  {https://ui.adsabs.harvard.edu/abs/2021MNRAS.502.4513W} {502, 4513}

\bibitem[\protect\citeauthoryear{{Zemp}, {Gnedin}, {Gnedin}  \&
  {Kravtsov}}{{Zemp} et~al.}{2011}]{zemp2011}
{Zemp} M.,  {Gnedin} O.~Y.,  {Gnedin} N.~Y.,   {Kravtsov} A.~V.,  2011, \mn@doi
  [\apjs] {10.1088/0067-0049/197/2/30}, \href
  {https://ui.adsabs.harvard.edu/abs/2011ApJS..197...30Z} {197, 30}

\bibitem[\protect\citeauthoryear{{del Pino}, {Fardal}, {van der Marel},
  {{\L}okas}, {Mateu}  \& {Sohn}}{{del Pino} et~al.}{2021}]{delpino2021}
{del Pino} A.,  {Fardal} M.~A.,  {van der Marel} R.~P.,  {{\L}okas} E.~L.,
  {Mateu} C.,   {Sohn} S.~T.,  2021, \mn@doi [\apj] {10.3847/1538-4357/abd5bf},
  \href {https://ui.adsabs.harvard.edu/abs/2021ApJ...908..244D} {908, 244}

\bibitem[\protect\citeauthoryear{{van Albada}}{{van
  Albada}}{1982}]{vanalbada1982}
{van Albada} T.~S.,  1982, \mn@doi [\mnras] {10.1093/mnras/201.4.939}, \href
  {https://ui.adsabs.harvard.edu/abs/1982MNRAS.201..939V} {201, 939}

\makeatother
\end{thebibliography}




\bsp	
\label{lastpage}
\end{document}